\documentclass[%
 reprint,
nofootinbib,
 amsmath,amssymb,
 aps,
]{revtex4-2}

\usepackage{graphicx}
\usepackage{dcolumn}
\usepackage{bm}
\usepackage{xcolor}
\newtheorem{definition}{Definition}

\begin{document}

\preprint{APS/123-QED}

\title{Is quantum advantage the right goal for quantum machine learning?}

\author{Maria Schuld}
\affiliation{Xanadu, Toronto, ON, M5G 2C8, Canada}
 \email{maria@xanadu.ai}

\author{Nathan Killoran}
\affiliation{Xanadu, Toronto, ON, M5G 2C8, Canada}

\date{\today}

\begin{abstract}
Machine learning is frequently listed among the most promising applications for quantum computing. This is in fact a curious choice: Today's machine learning algorithms are notoriously powerful in practice, but remain theoretically difficult to study. Quantum computing, in contrast, does not offer practical benchmarks on realistic scales, and theory is the main tool we have to judge whether it could become relevant for a problem. In this perspective we explain why it is so difficult to say something about the practical power of quantum computers for machine learning with the tools we are currently using. We argue that these challenges call for a critical debate on whether quantum advantage and the narrative of ``beating'' classical machine learning should continue to dominate the literature the way it does, and highlight examples for how other perspectives in existing research provide an important alternative to the focus on advantage.
\end{abstract}


\maketitle

The average number of papers on the arXiv's quantum section that relate to machine learning has increased from a handful of contributions per year in the early 2000s to a few papers \textit{per day} in 2021.\footnote{This observation is based on an arXiv API search in the \textit{quant-ph} category for papers using the term ``learn'' as substring in title or abstract. A similar trend is found for usage of the term ``neural network''.} A large share of this literature can be attributed to the field of quantum machine learning, which investigates how quantum computers can be used to solve machine learning problems \cite{wittek14, biamonte2017quantum, ciliberto2018quantum, dunjko2020non, schuld2021machine}, stemming from both conventional and ``quantum'' \cite{schatzki2021entangled, huang2021information} data. The dominant goal in quantum machine learning is to show that quantum computers, with their properties like entanglement and interference, offer advantages for machine learning tasks of practical relevance \cite{perdomo2018opportunities}. This question is particularly important to the emerging quantum technology industry, which has been driving the discipline right from the start \cite{neven08bin, amin16}, and which often names machine learning as one of the core application areas for quantum computers.

In this perspective, we want to put the goal of beating classical machine learning under critical scrutiny and argue that the scale of progress we seek may require at least a partial liberation from the ``tunnel vision of quantum advantage''. First, in Section \ref{sec:challenging} we explain why -- contrary to commercial expectations -- machine learning may turn out to be one of the hardest applications to show a practical quantum advantage for (see also Table \ref{tbl:comparison}): (a) machine learning is famous for notoriously powerful algorithms that set a challenging baseline for quantum algorithms, (b) the inputs to training algorithms are increasingly big and therefore hard to handle by early quantum computers, (c) the problems tend to stem from the human domain and are much messier than the tasks solved by standard quantum algorithms, (d) machine learning theory provides a shifting ground to work with, since past assumptions and intuition is currently being upheaved by \textit{deep learning}, and (e) we only have limited options to practically evaluate our methods with benchmarks. To state it in simple terms, quantum machine learning research is trying to beat large, high-performing algorithms for problems that are conceptually hard to study.

At the same time, the tools that quantum computing offers to think about advantages -- essentially, experiments on prototype quantum devices and over 30 years' worth of knowledge on provable asymptotic speedups -- are severely limited. Consequently, showing that quantum can beat classical machine learning may only be possible in highly abstract settings or on very small scales at this stage. Focusing on quantum advantage therefore means focusing only on a biased subset of models, datasets and theoretical approaches, namely the ones we can tackle under these difficult conditions -- a fact that we should discuss more critically. It is important to be clear that these challenges do not mean we should stop trying to figure out what quantum computers can offer for machine learning. But judging the value of research from  the limited lens of speedups could prevent important research areas from emerging, and in the worst case, it may even hinder the innovation needed to find use cases for quantum computers in the future. This is what we argue in Section \ref{sec:rethink}. 

However, if we decide to let go of the goal of beating classical machine learning for a moment, what other meaningful questions can we ask? In Section \ref{sec:links} we discuss how existing research already yields rich results without having to only look at quantum advantage. First, we use the discussion around \textit{quantum perceptrons} \cite{kapoor16, andrecut2002quantum, torrontegui2019unitary, killoran2019continuous, schuld2015simulating, espinosa15, benatti2019continuous, bondesan2020quantum} to motivate the question \textit{What are good building blocks for quantum models?}, where ``good'' can have much more diverse interpretations than speedups and beating benchmarks. Second, we explain how the connection between a large class of quantum machine learning models and kernel methods probes the important question \textit{How can we bridge quantum computing and classical learning theory to gain a better understanding of quantum machine learning?}, rather than only finding classically intractable quantum kernels. Third, we use the technique of computing gradients on a quantum computer as an example of a successful subfield that has not been purely driven by an advantage from the very beginning, but by the question \textit{How can we make quantum software ready for machine learning applications?}. We believe that all three ingredients -- the right quantum models to study, theoretical tools with which we can study them, and software solutions to scale experiments -- are important for a meaningful attempt to explore the benefit of quantum computing for machine learning in future. But, somewhat paradoxically, limiting what we deem worth researching in these and other areas by whether or not a paper can demonstrate that ``quantum is better'' may actually \textit{prevent} us from laying these much-needed foundations.

\begin{table*}
\def\arraystretch{1.5}

\begin{tabular}{p{2.5cm} p{7cm} p{0.5cm} p{7cm}}
\hline \hline
Property & Problems studied in quantum computing && Problems solved by machine learning\\
\hline
classical performance & \textbf{low} -- problems are carefully selected to be provably difficult for classical computers && \textbf{high} -- machine learning is applied on an industrial scale and many algorithms run in linear time in practice\\
size of inputs & \textbf{small} -- near-term algorithms are limited by small qubit numbers, while fault-tolerant algorithms usually take short bit strings  && \textbf{very large} -- may be millions of tensors with millions of entries each\\
problem structure & \textbf{very structured} -- often exhibiting a periodic structure that can be exploited by interference && \textbf{``messy''} -- problems are derived from the human or ``real-world'' domain and naturally complex to state and analyse\\
theoretical accessibility & \textbf{high} -- there is a large bias towards problems about which we can theoretically reason  && \textbf{shifting} -- theory is currently been re-built around the empirical success of deep learning \\
evaluating performance & \textbf{computational complexity} -- the dominant measure to assess the performance of an algorithm is asymptotic runtime scaling && \textbf{practical benchmarks} -- machine learning research puts a strong emphasis on empirical comparisons between methods \\
\hline \hline
\end{tabular}
\caption{Comparison of typical properties of problems studied in quantum computing versus problems solved by machine learning algorithms. Looking at this table, it is no surprise that quantum machine learning is a tough candidate for applications with a quantum advantage.}
\label{tbl:comparison}
\end{table*}

\section{Why machine learning is such a challenging problem}\label{sec:challenging}

To be more precise about why machine learning may be a challenging application for the state that quantum computing is in, we will have to become a bit more technical, and look at how a machine learning task can actually be formulated as a mathematical problem using the framework of empirical risk minimisation. This may contain familiar material to some readers, but will help us to make the argument of the following section more explicit.

\subsection{How to formalize learning}
Intuitively, learning is the acquisition of skills from examples\footnote{See \cite{chollet2019measure} for an interesting debate that is challenging modern machine learning, arguing that learning is the \textit{efficient} acquisition of skills from examples -- rather than just fitting massive models to massive datasets. } (some useful textbooks are \cite{bishop2006pattern, james2013introduction, goodfellow2016deep}). In machine learning, computers are the ``agents'' that learn, and examples are represented by \textit{data}. Skills can be as diverse as navigating a physical body in an environment, playing chess, generating artificial images, or translating languages. These situations have been captured by the famous distinction into supervised, unsupervised and reinforcement learning. It is a bit surprising at first that most of machine learning theory only focuses on supervised learning -- which is not so much a reflection of importance, but of the fact that supervision, or the provision of some information on the ``ground truth'', makes it easier to define what it means for a problem to be solved. At the same time, supervised learning does not have to deal with interaction between the learner and the data as is common in reinforcement learning.

A rather general version of a supervised learning problem can be stated as follows\footnote{There are many different ways to formalise the notion of learning, and as usual in science, there were strong trends in what machine learning research considered to be a ``relevant'' setting. For example, in the past \textit{learning from membership queries} \cite{servedio04}, where we can actively influence which data we are given, was often considered. A popular alternative flavour to supervised learning as we set it up here is \textit{PAC learning}, where learning translates to finding a model so that with a high probability, the loss of an input $x$ sampled from $p(x)$ has a loss smaller than a threshold.}:

\begin{definition}\textbf{(Supervised learning task)}
Consider a suitable data input domain $\mathcal{X}$ and a label domain $\mathcal{Y}$, as well as a probability distribution $p(x)$ over inputs $x \in \mathcal{X}$. We assume that there is some ground truth mapping $f^*: \mathcal{X} \to \mathcal{Y}$ of inputs to target labels. We are given a finite set of inputs sampled from $p(x)$, together with their target labels $ \mathcal{D} = \{(x^1, y^1),\dots, (x^M, y^M)\}$ with $(x, y) \in \mathcal{X} \otimes \mathcal{Y}$, as well as a loss $l: \mathcal{Y}\otimes \mathcal{Y} \to \mathbb{R}$ that tells us how well a label predicted by a function $f:\mathcal{X} \to \mathcal{Y}$ compares to the target label. The task is to find a model $f$ from a class of model functions $\mathcal{F}$ that minimizes the expected loss over the data distribution,
\begin{equation}
\hat{f} = \mathrm{argmin}_{f \in \mathcal{F}} \int_{\mathcal{X}} p(x)\;  l(f(x), f^*(x)) dx.
\label{eq:exp}
\end{equation}
\label{def:supervised}
\end{definition}

\begin{figure}
    \centering
    \includegraphics[width=0.27\textwidth]{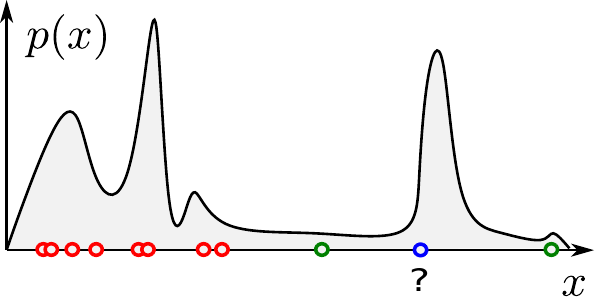}
    \caption{A central problem in machine learning is how to find a model that performs well with regards to a distribution $p(x)$ over datapoints $x$ if only a small set of samples from that distribution is given. In supervised learning, the data samples are labeled (red and green dots) and the goal is to label a new sample (blue dot). Especially in high dimensions, the samples will not be able to provide information on the entire data space (here indicated by the region of high density with no samples). Learning is only possible if the distribution, model and/or model selection strategy contains a lot of structure, which is not always easy to analyse theoretically.}
    \label{fig:data}
\end{figure}

For image recognition, which was one of the early success stories of modern machine learning, the inputs are numerical representations of images, and the labels could be a binary tag that indicates whether the image contains harmful content. The distribution $p(x)$ describes the probability with which we can expect to be given certain images in the problem, but is necessarily unknown in a real-life task. Instead, we are given a subset of example images drawn from this hypothetical distribution (see Fig.~\ref{fig:data}), as well as information on which contain harmful content ($y=1$) and which do not ($y=0$). A typical loss function is simply an indicator function\footnote{Note that optimization algorithms usually require continuous loss functions, and standard losses like least-squares can be understood as continuous surrogates for the indicator function.} 
\begin{equation}
    l(y, y') = \begin{cases} 1 \text{ if } y\neq y' \\ 0 \text{ else.} \end{cases}
\end{equation}
Minimising the expected loss over the data distribution is another way of saying that we want our model $f$ to do well with regards to the loss over all data we can expect to see. 

We can now pinpoint more precisely why machine learning applications are so hard to access from a theory point of view: in all but pathological examples, the probability distribution $p(x)$ as well as the target function $f^*$ in Definition~\ref{def:supervised} -- and hence an important part of Eq.~(\ref{eq:exp}) -- is unknown. Even if we could model it, the integral in Eq.~(\ref{eq:exp}) will be hard to compute for all but special cases. In other words, even a very basic formalization of machine learning translates to a mathematical problem that is usually unsolvable. 

\subsection{Solving the problem in practice}

Even though surprisingly few beginners to machine learning are aware of this correspondence, the standard approach of how to deal with this predicament is to solve a \textit{proxy problem} to Definition~\ref{def:supervised} and hope that it translates well to the original one. The proxy problem is known as \textit{empirical risk minimization}, and prescribes to evaluate the model performance using the finite set of data samples $\mathcal{D}$:
\begin{equation}
\hat{f}_{\rm emp} = \mathrm{argmin}_{f \in \mathcal{F}} 
\frac{1}{|\mathcal{D}|}\sum_{(x, y) \in \mathcal{D}} l(f(x), y).
\label{eq:empirical}
\end{equation}
Much of learning theory tries to find guarantees on how solving the empirical proxy will \textit{generalize} to the original problem, or how solutions found with a finite sample size perform on the original distribution.\footnote{It may not come as a surprise that the tools and terms of machine learning are largely borrowed from statistics.} 

The performance of a model on unseen data is usually measured on a test set of further data samples that have not been used for training, and most papers in the machine learning literature report the error on the test set by running benchmarks on famous datasets. While this sounds straightforward, getting high-quality results that do not depend on implementation details is hard. For quantum machine learning research, parts of which try to adopt the culture of benchmark comparisons, the current limitations of hardware size make it an even more challenging tool to use and interpret. 

In summary, while an important component of machine learning is optimization, its central aim is generalization, which is non-trivial to formalize and measure -- even more so when we want to add quantumness into the mix.

\subsection{Deep learning turns learning theory upside down}

Many of the standard tools in machine learning, such as cross-validation and regularization, are trying to fulfil the balancing act of not solving Eq.~(\ref{eq:empirical}) ``too well'': We want to use the information provided by the finite data sample, but we do not want to pick up its particularities (which may not be present if we were given a different data set  $\mathcal{D}$ sampled from $p(x)$). For example, if coincidentally all images in our data set that have a black pixel in one position are images with harmful content, we do not want to learn the spurious relation that if the pixel is black the image is harmful.

For the longest time, ``picking up too much information'' was thought to be identical to interpolating the training data perfectly well (i.e., getting a zero average loss over the data). Examples of mitigation strategies are to choose a simple function class $\mathcal{F}$, to add terms to the loss that penalise non-smooth models from that class, or to stop iterative optimization before it converges to a minimum. But since more than a decade, we are consistently getting empirical evidence that challenges this assumption: very large models can fit any function perfectly well, but still generalize beyond the data used for training -- even in the presence of noise. This phenomenon was first attributed to a kind of hierarchical model called a \textit{deep neural network}, but has been observed in other settings as well, and is now understood to be a main characteristic of the regime of so-called \textit{deep learning} \cite{zhang2021understanding, bartlett2021deep}. 

One of the most important goals in machine learning research today is to unite the evidence presented by deep learning with learning theory. This is a formidable challenge due to the mathematical structure of neural networks as long sequences of linear and nonlinear transformations, which make them unwieldy for mathematically modelling. Furthermore, it is by now largely uncontested that the algorithm with which neural networks are trained, as well as the data itself, plays a crucial role in the phenomena we observe in deep learning \cite{zhang2021understanding, soudry2018implicit, arora2019implicit}. A viable theory therefore cannot just make statements about the model class $\mathcal{F}$, but has to describe the solutions $\hat{f}$ to an optimization problem, as well as the data distribution $p$. This means that even the simplest of toy models has to capture many moving parts, each of which is already difficult to analyse in the first place. 

This ongoing revolution in machine learning theory-building, as well as the practical success of deep learning itself, obviously pose even more challenges for a theory of quantum machine learning, where we want to add quantum theory as another moving part. At the same time we have only little access to empirical results from ``just running the algorithm''. And even if few-qubit proof-of principle circuits can be simulated (or even run on real hardware), the learning regimes we are trying to understand are not observed on these small scales -- which means that we cannot say much about the behaviour that quantum models will exhibit on a realistic problem scale.

\section{A critical look at quantum advantage} \label{sec:rethink}

The previous section motivated why machine learning is a challenging problem to improve by quantum computers due to the good performance of existing algorithms, large inputs in many applications, the complex mathematical structure of the basic problems, and the little we know about why the best models perform so well, forcing us to gather evidence by benchmarks rather than guiding it by theory. 

In this section we will (after a short overview of the research field itself) motivate why in the context of machine learning the tools we currently use to investigate quantum advantage substantially limit and bias the statements we can make about the practical use of quantum computers. 

\subsection{Progress in quantum machine learning}

While sporadic papers at the intersection of quantum computing and machine learning were published since the 1990s \cite{kak95, bonnell97, ventura00, trugenberger02, schutzhold03, bonner03, servedio04, aimeur06, bshouty1998learning}, quantum machine learning -- here defined as research on how to use quantum computers for machine learning tasks from the classical or quantum domain -- only gained momentum around 2013 (see references in \cite{wittek14, schuld15qml}).  

Since then one can distinguish two popular approaches to quantum machine learning. In the first years, a common goal was to speed up existing machine learning algorithms by solving (sub)tasks such as matrix inversion \cite{wiebe12, rebentrost14, zlokapa2021quantum}, Gibbs sampling \cite{wiebe14b, denil11}, singular value estimation \cite{kerenidis2022quantum} or search \cite{kapoor16, aimeur13} on a quantum computer. Since this agenda is pretty much borrowed from the modus operandi of traditional quantum computing, it may not be surprising that the studies are firmly rooted in this parent discipline, and touch upon the intricacies of machine learning research only in the most basic strokes. 

The advent of near-term quantum computers led to a growing popularity of the second approach, which considers \textit{parametrized} or \textit{variational} quantum circuits as machine learning models  \cite{farhi18, schuld18cc, liu2018differentiable, benedetti2019parameterized}. In these proposals, training is done similarly to neural networks: gradient-descent-type algorithms iteratively find better physical parameters of the ``quantum model''. Central questions in this branch of research are what architectures to choose \cite{cong2019quantum, perez2020data}, how to compute gradients \cite{mitarai2018quantum, schuld2019evaluating}, as well as the trainability \cite{mcclean2018barren, holmes2021connecting}, expressivity \cite{wu2021expressivity, schuld2021effect} and generalization power \cite{huang2021power, abbas2021power, caro2021generalization, banchi2021generalization} of such models using insights from machine learning.

Apart from these two active fields of research there are many other contributions that try to formulate quantum versions of classical learning problems and analyse their scaling. For example we can ask how quantum data distributions change the sample complexity of learning \cite{servedio04, arunachalam2018optimal, huang2021information}, how classification problems change in a quantum setting \cite{sentis12, monras2017inductive,  banchi2021generalization}, how quantum agents learn from interacting with an environment \cite{paparo14, dunjko16}, or how quantum Ising models compare to Ising-based machine learning models such as Boltzmann machines \cite{amin16} or Hopfield networks \cite{rotondo2018open}.

\subsection{Quantum advantage} 

Almost all branches of quantum machine learning research have been heavily framed by the question of ``beating'' classical machine learning in some figure of merit, such as:
\begin{itemize}
   \item the asymptotic runtime of a particular machine learning algorithm, for example an optimiser used to solve the empirical risk minimisation problem in Eq.~(\ref{eq:empirical}) \cite{wiebe12, rebentrost14, zlokapa2021quantum, kerenidis2022quantum},
   \item whether or not a learning problem (like the one in Def.~\ref{def:supervised}) is efficiently solvable for a particular data distribution $p(x)$ \cite{liu2021rigorous, sweke2021quantum},
   \item the expressivity of a model class $\mathcal{F}$ \cite{du2020expressive, schuld2021effect}, 
   \item the number $M$ of samples needed to learn \cite{arunachalam2018optimal, huang2021quantum}, 
   \item average or worst-case generalization errors (which measure the difference between expected and empirical loss) \cite{caro2021generalization, abbas2021power, huang2021power},
   \item the structure of the optimisation landscape, giving us an idea of how easy it is to solve Eq.~(\ref{eq:empirical}) with gradient-based methods \cite{mcclean2018barren, holmes2021connecting}, or  
   \item the test error on some small-scale practical benchmark \cite{gili2022evaluating, selig2021case, schatzki2021entangled}. 
\end{itemize}

The positive publication bias known from other areas of science \cite{mlinaric2017dealing} is strongly prevalent in most areas of quantum machine learning, and a number of ``positive'' results have been put forward which either ``prove'' theoretically or ``show'' empirically that quantum computers are better at something. A few examples of the typical phrasing in abstracts and introductions are these: 
\begin{itemize}
    \item ``we establish a rigorous quantum speed-up for supervised classification'' \cite{liu2021rigorous}
    \item ``[w]e prove that [...] quantum machines can learn from exponentially fewer experiments than those required in conventional experiments.'' \cite{huang2021quantum}
    \item ``we prove that PQCs with a simple structure already outperform any classical neural network for generative tasks'' \cite{du2020expressive}
    \item ``for achieving accurate prediction on all inputs, we prove that exponential quantum advantage is possible'' \cite{huang2021information}
    \item ``[w]e show that our quantum-inspired generative models [..] generalize to unseen candidates with lower cost function values than any of the candidates seen by the classical solvers.'' \cite{alcazar2021enhancing}
    \item ``[o]ur simulation results show that our quantum-inspired models have up to a 68x enhancement in generating unseen [..] samples compared to GANs'' \cite{gili2022evaluating}
\end{itemize} 
Some areas, most notably the trainability of quantum models and the sample complexity in certain learning frameworks, actively discuss results that show which approaches do \textit{not} lead to advantages, or that could be problematic for practical quantum machine learning:
\begin{itemize}
    \item ``[w]e prove that for any input distribution [...], a classical ML model can provide accurate predictions on average by accessing [the quantum process generating data] a number of times comparable to the optimal quantum ML model'' \cite{huang2021information},
    \item ``despite concerns about gradient-based methods in classical deep neural networks [...], they are successful [... meanwhile, w]e show that for a large class of random circuits, the average value of the gradient of the objective function is zero'' \cite{mcclean2018barren},
    \item ``[o]ur main result is that quantum and classical sample complexity are in fact equal up to constant factors in both the PAC and agnostic models''  \cite{arunachalam2018optimal}.
\end{itemize}

If so much progress is being made in understanding quantum advantage, why do we think there is a problem? Will this research not eventually narrow down on areas where quantum computers could have a practical impact on machine learning applications -- or show by overwhelming evidence that the case is hopeless? We believe that there is a deeper structural issue: the tools we currently have in quantum computing are not sufficient to make meaningful statements about this question. Let us motivate this statement with a few points.

Proving exponential speedups for artificially constructed settings \cite{liu2018differentiable, sweke2021quantum}, on the one hand, is interesting from an academic point of view, but does not say much about possible quantum applications. In the language of machine learning, this approach picks problems that have a heavy bias in the data distribution $p(x)$ and/or the ground truth of the problem to what quantum computers solve well. Or as remarked in \cite{kubler2021inductive}, ``quantum machine learning models can offer speed-ups only if we manage to encode knowledge about the problem at hand into quantum circuits, while encoding the same bias into a classical model would be hard''. But as we explained in the previous section, the success of machine learning does not stem from solving a structured, well-understood problem and hand-coding it into the solution method. On the contrary, machine learning is famous for being agnostically applied to a range of problems for which we do not know the exact inductive bias that would suit the data. Furthermore, the principle of \textit{no free lunch} \cite{wolpert1996lack} in machine learning states that for any algorithm performing well on one problem there will always be another problem on which it does not perform well. Good performance on selected hand-crafted examples therefore does not tell us anything about quantum computers as general learning tools. Note that relaxing the need for exponential speedups can lead to provable advantages in more general settings,  (i.e., see \cite{paparo2014quantum, low2014quantum}), but it is currently questioned that such advantages will have an impact considering the overhead of error correction in fault-tolerant quantum computers \cite{babbush2021}.

The ``traditional approach'' to quantum machine learning mentioned in the previous section follows a different logic and looks for exponential speedups to widely applicable algorithms like support vector machines and neural nets \cite{rebentrost14, zlokapa2021quantum}. Since these algorithms evidently \textit{have} efficient runtimes, the goal is usually to reach sub-linear scaling. Here we find another issue, namely that we need extreme assumptions about data loading and read-out settings, and fair comparison to classical models have been challenged in the past \cite{aaronson15, tang2019quantum}.

Another potential issue of proving or disproving whether quantum machine learning ``works'' is the tendency of making statements about average or worst-case properties of extremely large model families, such as the class of models we can express as $f(x) = \mathrm{tr}\{\rho(x) M\}$ (where $\rho$ is a quantum state depending on $x$ and $M$ any observable), or models constructed from circuits sampled according to the Haar measure \cite{caro2021generalization, mcclean2018barren, kubler2021inductive}. Such statements do not preclude a more specific subclass of quantum models to have entirely different statistical properties. As a comparison, we may be able to prove that \textit{all} models we can express on a classical computer have certain average/worst-case properties for learning, which does not prevent \textit{specific} models like boosting or GANs to perform in an entirely different manner.

Empirical studies, on the other hand, tend to compare to very specific classical models on (necessarily) small datasets \cite{abbas2021power, schuld18cc}, and it is consequently hard to tell if advantages are due to the careful selection of the hyperparameters, benchmarks and comparisons, or if it is a structural observation. Small changes in the -- often ad-hoc designed -- architecture of the circuits can vary results significantly \cite{selig2021case}. Only few studies try to reproduce existing results \cite{franz2022uncovering}, or critically ask what measures we should apply \cite{gili2022evaluating} in our benchmarks other than borrowing concepts from classical machine learning. We also know very little about the scaling of empirical results to larger problem sizes, which will still be a challenge for experiments in years to come. 

In our view, the question about whether quantum computers can really play a role in identifying \textit{practical} machine learning applications is therefore still wide open, and unlikely to be decided by theoretical proofs or small-scale experiments. These tools should be considered more as a means to foster our understanding and test hypotheses in a well-defined setting. This is very relevant at the current state of quantum machine learning, where we observe an increasing resignation in informal conversations with colleagues and students as quantum machine learning fails to produce immediate commercial use-cases. The frequently repeated solution is to discard quantum computers for classical data processing \cite{kubler2021inductive}, and instead see the future of quantum machine learning in analysing data in the form of quantum states \cite{huang2021quantum, schatzki2021entangled}. But following the thoughts laid out in this section, we should ask ourselves if switching our attention to ``quantum data'' is subconsciously motivated by the hope that it suits our traditional proof techniques better, rather than providing a mature use-case. 

\section{Alternative research agendas}\label{sec:links}

Acknowledging the current difficulty of proposing quantum algorithms that improve the performance of machine learning does not mean that quantum machine learning research is at a dead end. Quite the contrary -- recent years have shown a lot of interesting and fruitful research areas which have grown our understanding of the intersection without focusing on advantages only. We want to illustrate this now with three examples.
The first two examples -- the search for a quantum perceptron and the link between quantum circuits and kernel methods -- shows how a research area can or has been framed from both an ``advantage'' and ``non-advantage'' perspective; either approach leads to different kinds of investigations which can mutually benefit from each other. The third example, the training of quantum circuits using gradients and automatic differentiation software, highlights an area that enabled quantum applications research without directly trying to improve classical algorithms in the first place.

\subsection{Quantum perceptrons or the search for building blocks of quantum models}

A perceptron \cite{rosenblatt58} is a simple function 
\begin{equation}
    f(\mathbf{x}) = \varphi(\mathbf{w}^T \mathbf{x} ),
    \label{eq:perc}
\end{equation} where $\mathbf{x}$ is an input vector, $\mathbf{w}$ a vector of trainable weights, and $\varphi$ a nonlinear scalar function. The perceptron has a long history that connects machine learning with biological models of the brain. It is the basic building block of neural networks, and hence most of the modern deep learning models used in practice today. Ways of constructing quantum versions of perceptrons have sparked the imagination of researchers since more than 25 years, and quantum machine learning consequently contains a huge variety of proposals (see for example \cite{kapoor16, benatti2019continuous, schuld2015simulating, andrecut2002quantum, lewenstein1994quantum, espinosa15, cao17, bondesan2020quantum, beer2020training, tacchino2019artificial}, to only mention a few).

Implicitly, quantum perceptrons are motivated by the success of classical perceptrons, and the desire to port this success over to the quantum domain. Depending on whether we want to prove a quantum advantage or not, very different study designs emerge. An advantage focus would require a comparison of quantum and classical versions with respect to runtime or performance in learning tasks. The design would have to focus on enabling this advantage (a feat that to our knowledge has not been convincingly performed yet). 

Shifting the motivation to other figures of merit allows us to shed a different light on the search for a quantum perceptron. As done in many studies, one could ask what the most natural equivalent of a non-linear activation function would be in quantum algorithms. But we could also try to find an efficiently trainable unit for quantum machine learning models that quantum hardware can easily implement. Other figures of merit are the simplicity of the model to allow theoretical investigations into training and generalisation behaviour, or whether it allows us to pinpoint ``non-classicality'' or ``quantumness'', so we can directly study its influence on learning. All these alternative figures of merit lead to very different design choices.

We want to remark in passing that without critical reflection the role of a universal building block has been filled by the ubiquitous Pauli rotations that we are so used to from quantum computing textbooks. But are we able to do better? Is there another ``unit'' that can provide a playground for theoretical insight and direct us towards the right practical implementations, such as the Ising model did for many-body physics \cite{taroni201590}, or linear models for deep learning \cite{bartlett2020benign, saxe2013exact}? Ironically, this change of perspective is not unlike the development of the perceptron itself: while researchers originally wanted to mimic a powerful concept for learning, namely the brain, porting it over to the computational domain required finding the right abstraction rather than emulating the original. Likewise, quantum researchers are trying to mimic the perceptron that has proven to be a powerful concept in classical machine learning, but it may turn out that rather than emulation, we ought to distill the crucial properties of this model to make it suitable for the quantum computing domain. 

\subsection{Quantum kernels as a bridge between quantum computing and learning theory}

\begin{figure}
    \centering
    \includegraphics[width=0.45\textwidth]{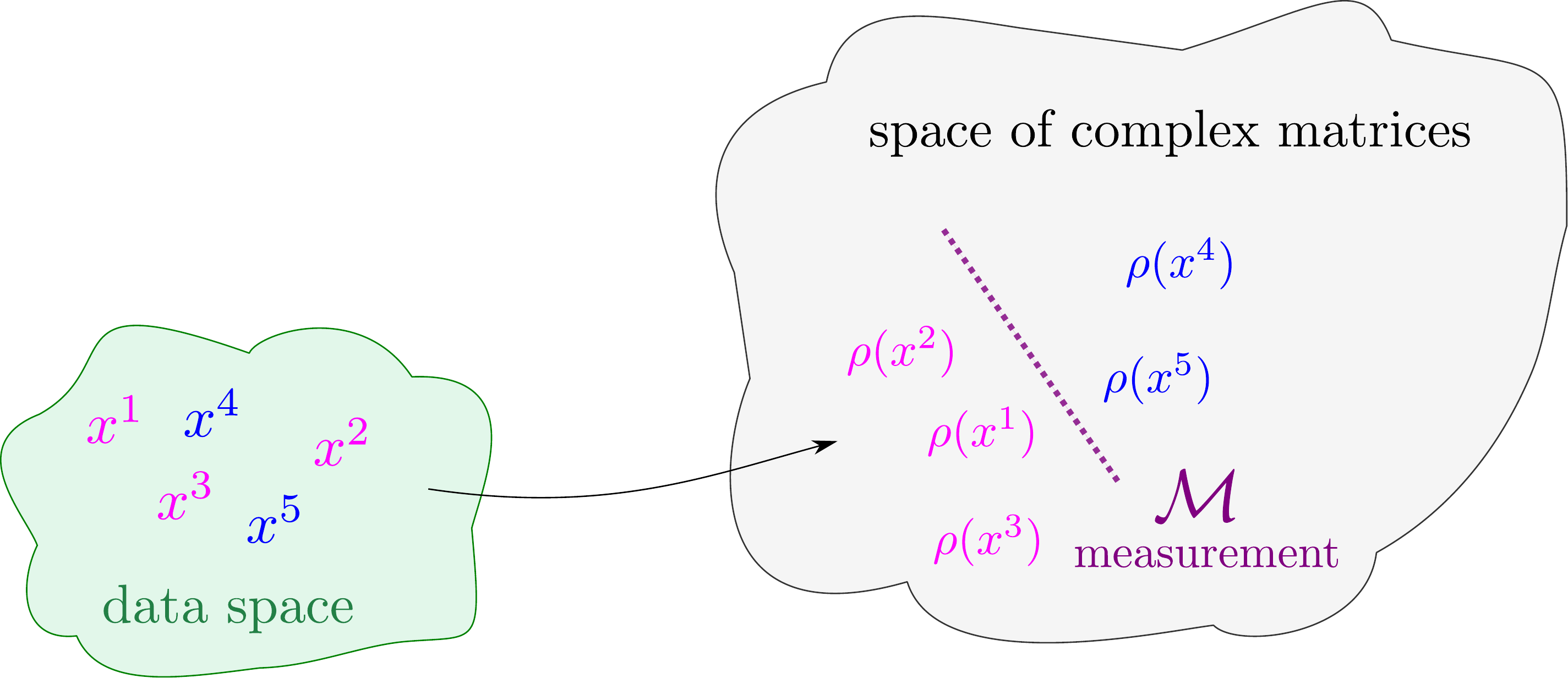}
    \caption{Many quantum circuits used as supervised machine learning models can be understood as mapping data to quantum states and then distinguishing these quantum states via hyperplanes defined by measurement observables. Such linear models in high-dimensional spaces are known as kernel methods in classical machine learning, and connect quantum machine learning to a rich set of tools to analyse optimization, learning and generalization.}
    \label{fig:linear}
\end{figure}

The second example we want to bring forward is that of quantum kernel methods. The research area of quantum machine learning grew out of the realization that data encoding is what machine learning researchers call a ``feature map'' \cite{rebentrost14, chatterjee2016generalized}, which means that many quantum circuits can be understood as a linear model in a feature space of the data \cite{schuld2019quantum, havlivcek2019supervised}. Again, part of this research area has been framed by (and used for) the search for quantum advantage \cite{huang2021power, havlivcek2019supervised, liu2021rigorous}. But there is a complementary angle: we can see this research area as an attempt to find formal connections between quantum and machine learning theory, connections which help us to apply results from one field to the other (see also \cite{schuld2021quantum}). We want to briefly introduce the basic concepts of quantum kernel research (see Fig.~\ref{fig:linear}) to compare these two angles in more detail.   

In a nutshell, quantum kernel research is based on the insight that if we encode a data input $x \in \mathcal{X}$ into a quantum state $\rho(x)$ (for example via a quantum state preparation routine), the expectation of an observable $\mathcal{M}$ can be interpreted as a machine learning model of the form 
\begin{equation}
    f(x) = \rm{tr}\{\rho(x) \mathcal{M}\}.
    \label{eq:qmodel}
\end{equation}
Realising that the trace is an inner product (known as the Hilbert-Schmidt inner product) in the space of complex-valued matrices, and that $\rho(x)$ maps the input $x$ into this space, we can state that the ``quantum model'' in Eq.~(\ref{eq:qmodel}) is a \textit{linear model} of the form
\begin{equation}
    f(x) = \langle \phi(x), w \rangle_{\mathcal{H}},
    \label{eq:linear}
\end{equation}
where $\phi(x)$ is a \textit{feature map} from the data space to a feature space $\mathcal{H}$, $w$ a weight vector, and $\langle \cdot, \cdot \rangle$ the inner product in $\mathcal{H}$. Most often, $\mathcal{H}$ is simply $\mathbb{R}^N$. The weight vector then contains trainable parameters and defines a linear hyperplane that can be used to separate classes of data in a supervised learning problem. Likewise, in many variational quantum models $\mathcal{M} = \mathcal{M}(\theta)$ from Eq.~(\ref{eq:qmodel}) is trainable: by optimising a parametrized circuit before a fixed measurement, we effectively choose a measurement basis (and hence the discriminating hyperplane) via optimization. 

This innocent link has immense consequences. Linear models in high-dimensional spaces are the core of one of the richest corners of machine learning theory, namely kernel theory, which we can now use to understand ``quantum models'' \cite{schuld2021quantum}. For example, kernel theory tells us that quantum models of the form in Eq.~(\ref{eq:qmodel}) can be rewritten as a linear combination of the ``distances'' between quantum states encoding the training data points $x^m$ and the quantum state encoding the input $x$ we seek to classify,
\begin{equation}
    f(x) = \sum_{m=1}^M a_m \rm{tr}\{\rho(x) \rho(x^m)\}, \;\; x_m \in \mathcal{D}, a_m \in \mathbb{R}.
    \label{eq:emp_kernel}
\end{equation}
Instead of learning the parameters $\theta$ in a variational circuit, we can learn the coefficients $a_m$ and only need the quantum computer to evaluate the trace term (which for pure states reduces to the overlap $|\langle \psi(x)| \psi(x^m)\rangle |^2$). Furthermore, we are guaranteed that the optimal coefficients $a_m$ construct a model that is also the global minimum of the empirical risk minimization problem in Eq.~(\ref{eq:empirical}). In other words, while Eq.~(\ref{eq:emp_kernel}) may define a smaller function class compared to Eq.~(\ref{eq:qmodel}), it still contains the solution we want to find. If the loss used to compare predictions with target labels is convex, the entire optimization problem is convex and hence conceptually simple to analyze. It also guarantees that we can find the optimal solution. 

Understanding quantum computers as ``kernel evaluators'' can tell us something about quantum advantage. In situations where this link holds, potential speedups have to be located in the evaluation of the kernel $\rm{tr}\{\rho(x), \rho(x')\}$ \cite{huang2021power}. We can investigate kernels based on circuits that are believed to be classically intractable \cite{havlivcek2019supervised}, and prove end-to-end quantum advantages for very specific learning problems \cite{liu2021rigorous}. At the same time, the necessity to estimate the value of the kernel function using finite shots introduces an overhead \cite{gentinetta2022complexity, peters2021machine}. While this research is certainly valuable, a classically intractable kernel that is \textit{useful} for practical machine learning tasks has yet to be found.

On the other hand, we can view quantum kernel theory purely as a tool for theory-building. For example, the link connects quantum circuits to linear representations of neural networks such as neural tangent kernels \cite{rahimi2007random} and random Fourier features \cite{jacot2018neural}, which are central to current investigations of deep learning -- a fact that has been explored in a series of recent papers \cite{shirai2021quantum, nakaji2021quantum, liu2021representation}. The theory of kernel methods also allows us to study generalization by making statements about the margin between the data-encoding states $\rho(x)$ for two different classes of data \cite{banchi2021generalization} or the regularization properties of a model \cite{steinwart2008}. Finally, it allows us to port over insights from quantum state discrimination as to what optimal decision boundaries are \cite{lloyd2020quantum}. None of these studies directly tries to answer the question of whether or not quantum computers could be superior for learning, and leads to very different kinds of results.

There are many similar points of contact between quantum and machine learning theory, such as the interpretation of quantum measurements as samples from a generative model \cite{cheng2018information}, or the proximity of quantum computers to machine learning models inspired by many-body-physics \cite{amin16}, and the usefulness of neural networks in representing quantum states \cite{carleo2017solving}.

\subsection{Quantum gradients and making quantum software ready for machine learning applications}

The last example highlights an area of research that massively increased our capability of performing experiments and building software around quantum machine learning without having quantum advantages as an immediate goal. It is the study of gradients of quantum computations, and how to retrieve them from performing other, efficient, quantum computations. In fact, it is well known that so-called parameter-shift rules \cite{mitarai2018quantum, schuld2019evaluating} put forward for this task are \textit{less} efficient than classical backpropagation, since they require a full model estimation per model parameter (while each estimation requires many shots or runs of the model circuit).

Historically, the dominant representation of quantum computations involved static algorithms that were hand-designed by expert theorists to maximally leverage coherent effects. 
More recently, there has been growing recognition that adding free parameters to quantum circuits allows them to represent an entire family of functions, while retaining the unique coherence properties that make quantum algorithms distinct \cite{cerezo2021variational}. 
The best value of these parameters for a particular task can then be determined variationally.
This expansion makes it easier for researchers to quickly test out new ideas and discover new quantum algorithms — as evidenced by the recent explosion of works on variational quantum circuits — but comes with the caveat that such classes of circuits may be harder to pin down theoretically (compared to, for example, kernel methods discussed in the previous section). 
Notably, this dichotomy mirrors the present-day situation in deep learning.

In the variational framework, a quantum circuit implements a function of the form\footnote{In some settings, the input data $x$ might be omitted. In these cases, since no data is present, the problem is more formally an optimization problem and not a machine learning problem. Because the tools employed are exactly the same, these distinctions are often conflated.}
\begin{equation}
    f(x, \theta) = \rm{tr}\{\rho(x, \theta) \mathcal{M}(\theta)\},
\end{equation}
where in contrast to Eq.~(\ref{eq:linear}) we included the free parameters $\theta$ in the measurement, and also allowed for a trainable state $\rho(x, \theta)$.
Typically, the free parameters correspond to rotation angles of gates in a quantum circuit.
This presents us with a new task: given a parametrized circuit, how should we adjust the parameter values to ``train'' the circuit to minimize some loss function $l$ that measures the quality of $f(x, \theta)$?

\begin{figure}
    \centering
    \includegraphics[width=0.45\textwidth]{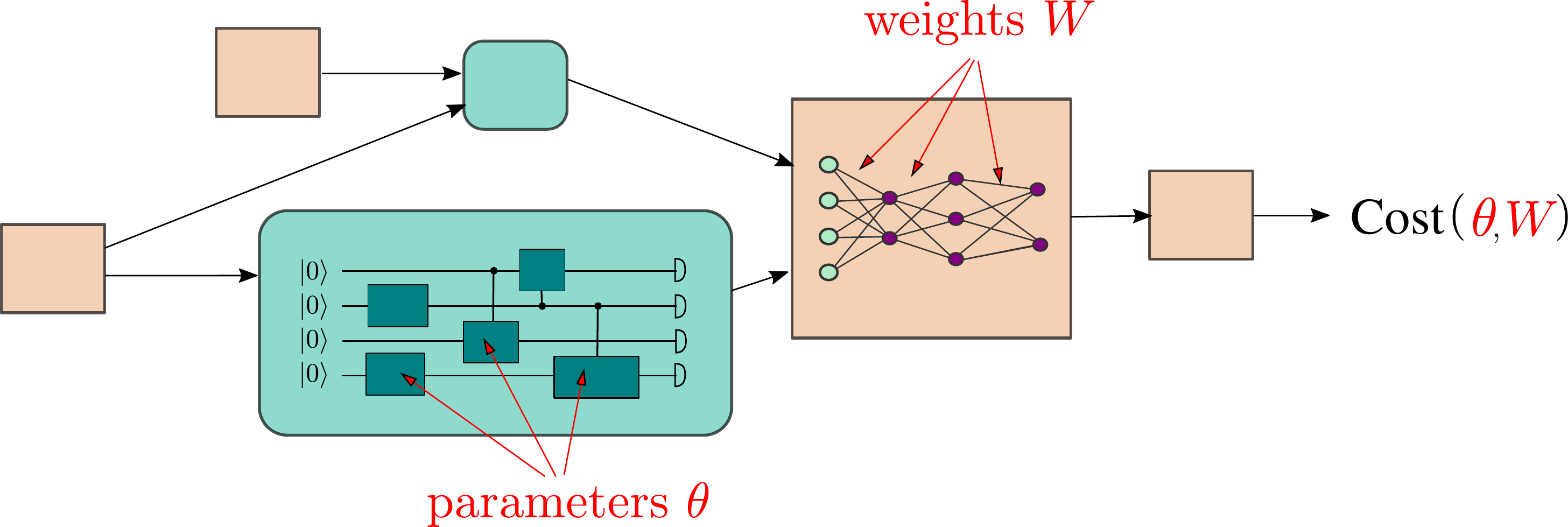}
    \caption{Parametrized quantum circuits can be trained as parts of larger machine learning pipelines by making use of automatic differentiation and the fact that we know in many settings how to estimate analytic gradients of cost functions with respect to circuit parameters.}
    \label{fig:train}
\end{figure}

While many options are available for training, there are very intriguing links with the workhorse algorithm used to train deep learning models: gradient descent.
In gradient descent, we optimize a loss function by computing its gradient with respect to the free parameters, and iteratively updating the parameters in the direction of the gradient.
From the chain rule, we must therefore determine the gradient of the model function, $\nabla_\theta f(x, \theta),$ with respect to the circuit's free parameters $\theta$.
Modern software tools like TensorFlow \cite{abadi2016tensorflow} or PyTorch \cite{paszke2019pytorch} largely automate the gradient computation of deep learning models using the backpropagation algorithm \cite{rumelhart1986learning}.
These libraries even let a user optimize custom functions — such as an expectation value produced from calling quantum computing hardware — provided the user also supplies the gradient of this function.

Drawing on insights originally developed from quantum optimal control \cite{li2017hybrid}, it turns out to be remarkably simple to compute the gradients of (many) quantum circuits. Using a technique now known as the parameter-shift rule \cite{mitarai2018quantum, schuld2019evaluating}, we can evaluate the derivatives $\frac{\partial f}{\partial \theta_i}$ of a parametrized circuit\footnote{In this case, the parameters should appear as the rotation angles of gates.} — and hence the gradient as well — by running the same circuit with parameter $\theta_i$ shifted forward and backward by a fixed amount,
\begin{equation}
    \frac{\partial f}{\partial \theta_i} = \frac{f(x, \theta + s\hat{e}_i)-f(x, \theta - s\hat{e}_i)}{2\sin(s)}.
\end{equation}
This technique, which has since been generalized to more and more cases \cite{crooks2019gradients, banchi2021measuring, kottmann2021feasible, vidal2018calculus, mari2021estimating, izmaylov2021analytic, kyriienko2021generalized, wierichs2021general}, has a similar form to the numerical finite-difference approximator, but in fact provides an analytically exact expression\footnote{In the case of expectation values obtained with a finite number of shots, the parameter-shift rule remains an unbiased estimator for the analytic gradient.} for any shift value $s\neq 0,\pi$.
And although it does not match the efficiency of backpropagation\footnote{In backpropagation, or automatic differentiation \textit{through} models, we typically only step through the algorithm once forward and once backward —reusing computations cached from the forward pass — to get the entire gradient vector. In contrast, each entry in the gradient of a quantum circuit requires additional circuit evaluations on top of the evaluation of $f(x,\theta)$.}, the simplicity of the parameter-shift rule makes it a very hardware-friendly mechanism for computing quantum circuit gradients.

Parameter-shift rules, quantum gradients and the resulting surge in quantum software for automatic differentiation are a prime example of research that strives by \textit{enabling} quantum machine learning applications, rather than demanding superiority of quantum algorithms. Armed with the ability to evaluate quantum models and compute their gradients, we can directly ``plug and play'' with existing deep learning tools and train quantum circuits the same way as we train neural networks. 
We can connect differentiable quantum subroutines into larger hybrid quantum-classical models and train the whole pipeline end-to-end using any of the specialized gradient-based optimizers developed in deep learning, such as Momentum or Adam \cite{kingma2014adam} (see Fig.~\ref{fig:train}). 
Finally, the links unveiled through the study of quantum gradients and training quantum models open up a rich opportunity for cross-pollination of ideas between quantum computing and deep learning.
For example, we have already seen the arrival of ``quantum-aware'' optimizers \cite{stokes2020quantum, kubler2020adaptive,  arrasmith2020operator, ostaszewski2021structure} which tweak ideas from deep learning to make them more native to the quantum setting.
On the theory side, we can leverage the latest (admittedly, still evolving) theoretical insights on optimization landscapes and generalizations coming from deep learning, and potentially adapt them to better understand phenomena such as barren plateaus \cite{mcclean2018barren}.
As our understanding increases, ideas and techniques from quantum computing can even find their way back into deep learning. A recent example is the use of tensor-network-based models in place of standard neural networks \cite{stoudenmire2018learning}.

\section{Moving forward}\label{sec:conclusion}

This perspective advocated a shift in the research agenda of quantum machine learning away from investing all our resources into the notion of ``beating'' classical algorithms. Sections \ref{sec:challenging} and \ref{sec:rethink} tried to motivate such a shift by arguing that the goal of showing quantum advantages forces us to limit our analytical focus to the very few problems we can actually study in a setting as complex as machine learning, while Section~\ref{sec:links} showcased existing areas framed by alternative research questions. Until quantum computers become available to do large-scale benchmarks, asking more fundamental questions may be a very good use of our time, but require a bit of courage to withstand the narrative of trying to find the billion-dollar quantum ``supremacy'', or to resist catchy expressions like ``the power of deep quantum neural networks''. 

A paradigm shift is never easy, and will require the community to make subtle but crucial adjustments, for example to the way that supervisors guide students, how science journalists portray the topic, how companies formulate their deliverables, and how reviewers judge publication-worthiness. However, in the end this may be exactly what is needed to push quantum machine learning research to the level that leads to future industrial-scale applications.

\bibliography{lit}

\begin{thebibliography}{123}%
\makeatletter
\providecommand \@ifxundefined [1]{%
 \@ifx{#1\undefined}
}%
\providecommand \@ifnum [1]{%
 \ifnum #1\expandafter \@firstoftwo
 \else \expandafter \@secondoftwo
 \fi
}%
\providecommand \@ifx [1]{%
 \ifx #1\expandafter \@firstoftwo
 \else \expandafter \@secondoftwo
 \fi
}%
\providecommand \natexlab [1]{#1}%
\providecommand \enquote  [1]{``#1''}%
\providecommand \bibnamefont  [1]{#1}%
\providecommand \bibfnamefont [1]{#1}%
\providecommand \citenamefont [1]{#1}%
\providecommand \href@noop [0]{\@secondoftwo}%
\providecommand \href [0]{\begingroup \@sanitize@url \@href}%
\providecommand \@href[1]{\@@startlink{#1}\@@href}%
\providecommand \@@href[1]{\endgroup#1\@@endlink}%
\providecommand \@sanitize@url [0]{\catcode `\\12\catcode `\$12\catcode
  `\&12\catcode `\#12\catcode `\^12\catcode `\_12\catcode `\%12\relax}%
\providecommand \@@startlink[1]{}%
\providecommand \@@endlink[0]{}%
\providecommand \url  [0]{\begingroup\@sanitize@url \@url }%
\providecommand \@url [1]{\endgroup\@href {#1}{\urlprefix }}%
\providecommand \urlprefix  [0]{URL }%
\providecommand \Eprint [0]{\href }%
\providecommand \doibase [0]{https://doi.org/}%
\providecommand \selectlanguage [0]{\@gobble}%
\providecommand \bibinfo  [0]{\@secondoftwo}%
\providecommand \bibfield  [0]{\@secondoftwo}%
\providecommand \translation [1]{[#1]}%
\providecommand \BibitemOpen [0]{}%
\providecommand \bibitemStop [0]{}%
\providecommand \bibitemNoStop [0]{.\EOS\space}%
\providecommand \EOS [0]{\spacefactor3000\relax}%
\providecommand \BibitemShut  [1]{\csname bibitem#1\endcsname}%
\let\auto@bib@innerbib\@empty
\bibitem [{\citenamefont {Wittek}(2014)}]{wittek14}%
  \BibitemOpen
  \bibfield  {author} {\bibinfo {author} {\bibfnamefont {P.}~\bibnamefont
  {Wittek}},\ }\href@noop {} {\emph {\bibinfo {title} {Quantum machine
  learning: {W}hat quantum computing means to data mining}}}\ (\bibinfo
  {publisher} {Academic Press},\ \bibinfo {year} {2014})\BibitemShut {NoStop}%
\bibitem [{\citenamefont {Biamonte}\ \emph {et~al.}(2017)\citenamefont
  {Biamonte}, \citenamefont {Wittek}, \citenamefont {Pancotti}, \citenamefont
  {Rebentrost}, \citenamefont {Wiebe},\ and\ \citenamefont
  {Lloyd}}]{biamonte2017quantum}%
  \BibitemOpen
  \bibfield  {author} {\bibinfo {author} {\bibfnamefont {J.}~\bibnamefont
  {Biamonte}}, \bibinfo {author} {\bibfnamefont {P.}~\bibnamefont {Wittek}},
  \bibinfo {author} {\bibfnamefont {N.}~\bibnamefont {Pancotti}}, \bibinfo
  {author} {\bibfnamefont {P.}~\bibnamefont {Rebentrost}}, \bibinfo {author}
  {\bibfnamefont {N.}~\bibnamefont {Wiebe}},\ and\ \bibinfo {author}
  {\bibfnamefont {S.}~\bibnamefont {Lloyd}},\ }\bibfield  {title} {\bibinfo
  {title} {Quantum machine learning},\ }\href@noop {} {\bibfield  {journal}
  {\bibinfo  {journal} {Nature}\ }\textbf {\bibinfo {volume} {549}},\ \bibinfo
  {pages} {195} (\bibinfo {year} {2017})}\BibitemShut {NoStop}%
\bibitem [{\citenamefont {Ciliberto}\ \emph {et~al.}(2018)\citenamefont
  {Ciliberto}, \citenamefont {Herbster}, \citenamefont {Ialongo}, \citenamefont
  {Pontil}, \citenamefont {Rocchetto}, \citenamefont {Severini},\ and\
  \citenamefont {Wossnig}}]{ciliberto2018quantum}%
  \BibitemOpen
  \bibfield  {author} {\bibinfo {author} {\bibfnamefont {C.}~\bibnamefont
  {Ciliberto}}, \bibinfo {author} {\bibfnamefont {M.}~\bibnamefont {Herbster}},
  \bibinfo {author} {\bibfnamefont {A.~D.}\ \bibnamefont {Ialongo}}, \bibinfo
  {author} {\bibfnamefont {M.}~\bibnamefont {Pontil}}, \bibinfo {author}
  {\bibfnamefont {A.}~\bibnamefont {Rocchetto}}, \bibinfo {author}
  {\bibfnamefont {S.}~\bibnamefont {Severini}},\ and\ \bibinfo {author}
  {\bibfnamefont {L.}~\bibnamefont {Wossnig}},\ }\bibfield  {title} {\bibinfo
  {title} {Quantum machine learning: a classical perspective},\ }\href@noop {}
  {\bibfield  {journal} {\bibinfo  {journal} {Proceedings of the Royal Society
  A: Mathematical, Physical and Engineering Sciences}\ }\textbf {\bibinfo
  {volume} {474}},\ \bibinfo {pages} {20170551} (\bibinfo {year}
  {2018})}\BibitemShut {NoStop}%
\bibitem [{\citenamefont {Dunjko}\ and\ \citenamefont
  {Wittek}(2020)}]{dunjko2020non}%
  \BibitemOpen
  \bibfield  {author} {\bibinfo {author} {\bibfnamefont {V.}~\bibnamefont
  {Dunjko}}\ and\ \bibinfo {author} {\bibfnamefont {P.}~\bibnamefont
  {Wittek}},\ }\bibfield  {title} {\bibinfo {title} {A non-review of quantum
  machine learning: trends and explorations},\ }\href@noop {} {\bibfield
  {journal} {\bibinfo  {journal} {Quantum Views}\ }\textbf {\bibinfo {volume}
  {4}},\ \bibinfo {pages} {32} (\bibinfo {year} {2020})}\BibitemShut {NoStop}%
\bibitem [{\citenamefont {Schuld}\ and\ \citenamefont
  {Petruccione}(2021)}]{schuld2021machine}%
  \BibitemOpen
  \bibfield  {author} {\bibinfo {author} {\bibfnamefont {M.}~\bibnamefont
  {Schuld}}\ and\ \bibinfo {author} {\bibfnamefont {F.}~\bibnamefont
  {Petruccione}},\ }\href@noop {} {\emph {\bibinfo {title} {Machine Learning
  with Quantum Computers}}}\ (\bibinfo  {publisher} {Springer},\ \bibinfo
  {year} {2021})\BibitemShut {NoStop}%
\bibitem [{\citenamefont {Schatzki}\ \emph {et~al.}(2021)\citenamefont
  {Schatzki}, \citenamefont {Arrasmith}, \citenamefont {Coles},\ and\
  \citenamefont {Cerezo}}]{schatzki2021entangled}%
  \BibitemOpen
  \bibfield  {author} {\bibinfo {author} {\bibfnamefont {L.}~\bibnamefont
  {Schatzki}}, \bibinfo {author} {\bibfnamefont {A.}~\bibnamefont {Arrasmith}},
  \bibinfo {author} {\bibfnamefont {P.~J.}\ \bibnamefont {Coles}},\ and\
  \bibinfo {author} {\bibfnamefont {M.}~\bibnamefont {Cerezo}},\ }\bibfield
  {title} {\bibinfo {title} {Entangled datasets for quantum machine learning},\
  }\href@noop {} {\bibfield  {journal} {\bibinfo  {journal} {arXiv preprint
  arXiv:2109.03400}\ } (\bibinfo {year} {2021})}\BibitemShut {NoStop}%
\bibitem [{\citenamefont {Huang}\ \emph
  {et~al.}(2021{\natexlab{a}})\citenamefont {Huang}, \citenamefont {Kueng},\
  and\ \citenamefont {Preskill}}]{huang2021information}%
  \BibitemOpen
  \bibfield  {author} {\bibinfo {author} {\bibfnamefont {H.-Y.}\ \bibnamefont
  {Huang}}, \bibinfo {author} {\bibfnamefont {R.}~\bibnamefont {Kueng}},\ and\
  \bibinfo {author} {\bibfnamefont {J.}~\bibnamefont {Preskill}},\ }\bibfield
  {title} {\bibinfo {title} {Information-theoretic bounds on quantum advantage
  in machine learning},\ }\href@noop {} {\bibfield  {journal} {\bibinfo
  {journal} {arXiv preprint arXiv:2101.02464}\ } (\bibinfo {year}
  {2021}{\natexlab{a}})}\BibitemShut {NoStop}%
\bibitem [{\citenamefont {Perdomo-Ortiz}\ \emph {et~al.}(2018)\citenamefont
  {Perdomo-Ortiz}, \citenamefont {Benedetti}, \citenamefont
  {Realpe-G{\'o}mez},\ and\ \citenamefont {Biswas}}]{perdomo2018opportunities}%
  \BibitemOpen
  \bibfield  {author} {\bibinfo {author} {\bibfnamefont {A.}~\bibnamefont
  {Perdomo-Ortiz}}, \bibinfo {author} {\bibfnamefont {M.}~\bibnamefont
  {Benedetti}}, \bibinfo {author} {\bibfnamefont {J.}~\bibnamefont
  {Realpe-G{\'o}mez}},\ and\ \bibinfo {author} {\bibfnamefont {R.}~\bibnamefont
  {Biswas}},\ }\bibfield  {title} {\bibinfo {title} {Opportunities and
  challenges for quantum-assisted machine learning in near-term quantum
  computers},\ }\href@noop {} {\bibfield  {journal} {\bibinfo  {journal}
  {Quantum Science and Technology}\ }\textbf {\bibinfo {volume} {3}},\ \bibinfo
  {pages} {030502} (\bibinfo {year} {2018})}\BibitemShut {NoStop}%
\bibitem [{\citenamefont {Neven}\ \emph {et~al.}(2008)\citenamefont {Neven},
  \citenamefont {Denchev}, \citenamefont {Rose},\ and\ \citenamefont
  {Macready}}]{neven08bin}%
  \BibitemOpen
  \bibfield  {author} {\bibinfo {author} {\bibfnamefont {H.}~\bibnamefont
  {Neven}}, \bibinfo {author} {\bibfnamefont {V.~S.}\ \bibnamefont {Denchev}},
  \bibinfo {author} {\bibfnamefont {G.}~\bibnamefont {Rose}},\ and\ \bibinfo
  {author} {\bibfnamefont {W.~G.}\ \bibnamefont {Macready}},\ }\bibfield
  {title} {\bibinfo {title} {Training a binary classifier with the quantum
  adiabatic algorithm},\ }\href@noop {} {\bibfield  {journal} {\bibinfo
  {journal} {arXiv preprint arXiv:0811.0416}\ } (\bibinfo {year}
  {2008})}\BibitemShut {NoStop}%
\bibitem [{\citenamefont {Amin}\ \emph {et~al.}(2018)\citenamefont {Amin},
  \citenamefont {Andriyash}, \citenamefont {Rolfe}, \citenamefont
  {Kulchytskyy},\ and\ \citenamefont {Melko}}]{amin16}%
  \BibitemOpen
  \bibfield  {author} {\bibinfo {author} {\bibfnamefont {M.~H.}\ \bibnamefont
  {Amin}}, \bibinfo {author} {\bibfnamefont {E.}~\bibnamefont {Andriyash}},
  \bibinfo {author} {\bibfnamefont {J.}~\bibnamefont {Rolfe}}, \bibinfo
  {author} {\bibfnamefont {B.}~\bibnamefont {Kulchytskyy}},\ and\ \bibinfo
  {author} {\bibfnamefont {R.}~\bibnamefont {Melko}},\ }\bibfield  {title}
  {\bibinfo {title} {Quantum {B}oltzmann machine},\ }\href@noop {} {\bibfield
  {journal} {\bibinfo  {journal} {Physical {R}eview {X}}\ }\textbf {\bibinfo
  {volume} {8}},\ \bibinfo {pages} {021050} (\bibinfo {year}
  {2018})}\BibitemShut {NoStop}%
\bibitem [{\citenamefont {Kapoor}\ \emph {et~al.}(2016)\citenamefont {Kapoor},
  \citenamefont {Wiebe},\ and\ \citenamefont {Svore}}]{kapoor16}%
  \BibitemOpen
  \bibfield  {author} {\bibinfo {author} {\bibfnamefont {A.}~\bibnamefont
  {Kapoor}}, \bibinfo {author} {\bibfnamefont {N.}~\bibnamefont {Wiebe}},\ and\
  \bibinfo {author} {\bibfnamefont {K.}~\bibnamefont {Svore}},\ }\bibfield
  {title} {\bibinfo {title} {Quantum perceptron models},\ }in\ \href@noop {}
  {\emph {\bibinfo {booktitle} {Advances In Neural Information Processing
  Systems}}}\ (\bibinfo {year} {2016})\ pp.\ \bibinfo {pages}
  {3999--4007}\BibitemShut {NoStop}%
\bibitem [{\citenamefont {Andrecut}\ and\ \citenamefont
  {Ali}(2002)}]{andrecut2002quantum}%
  \BibitemOpen
  \bibfield  {author} {\bibinfo {author} {\bibfnamefont {M.}~\bibnamefont
  {Andrecut}}\ and\ \bibinfo {author} {\bibfnamefont {M.}~\bibnamefont {Ali}},\
  }\bibfield  {title} {\bibinfo {title} {A quantum perceptron},\ }\href@noop {}
  {\bibfield  {journal} {\bibinfo  {journal} {International Journal of Modern
  Physics B}\ }\textbf {\bibinfo {volume} {16}},\ \bibinfo {pages} {639}
  (\bibinfo {year} {2002})}\BibitemShut {NoStop}%
\bibitem [{\citenamefont {Torrontegui}\ and\ \citenamefont
  {Garc{\'\i}a-Ripoll}(2019)}]{torrontegui2019unitary}%
  \BibitemOpen
  \bibfield  {author} {\bibinfo {author} {\bibfnamefont {E.}~\bibnamefont
  {Torrontegui}}\ and\ \bibinfo {author} {\bibfnamefont {J.~J.}\ \bibnamefont
  {Garc{\'\i}a-Ripoll}},\ }\bibfield  {title} {\bibinfo {title} {Unitary
  quantum perceptron as efficient universal approximator},\ }\href@noop {}
  {\bibfield  {journal} {\bibinfo  {journal} {EPL (Europhysics Letters)}\
  }\textbf {\bibinfo {volume} {125}},\ \bibinfo {pages} {30004} (\bibinfo
  {year} {2019})}\BibitemShut {NoStop}%
\bibitem [{\citenamefont {Killoran}\ \emph {et~al.}(2019)\citenamefont
  {Killoran}, \citenamefont {Bromley}, \citenamefont {Arrazola}, \citenamefont
  {Schuld}, \citenamefont {Quesada},\ and\ \citenamefont
  {Lloyd}}]{killoran2019continuous}%
  \BibitemOpen
  \bibfield  {author} {\bibinfo {author} {\bibfnamefont {N.}~\bibnamefont
  {Killoran}}, \bibinfo {author} {\bibfnamefont {T.~R.}\ \bibnamefont
  {Bromley}}, \bibinfo {author} {\bibfnamefont {J.~M.}\ \bibnamefont
  {Arrazola}}, \bibinfo {author} {\bibfnamefont {M.}~\bibnamefont {Schuld}},
  \bibinfo {author} {\bibfnamefont {N.}~\bibnamefont {Quesada}},\ and\ \bibinfo
  {author} {\bibfnamefont {S.}~\bibnamefont {Lloyd}},\ }\bibfield  {title}
  {\bibinfo {title} {Continuous-variable quantum neural networks},\ }\href@noop
  {} {\bibfield  {journal} {\bibinfo  {journal} {Physical Review Research}\
  }\textbf {\bibinfo {volume} {1}},\ \bibinfo {pages} {033063} (\bibinfo {year}
  {2019})}\BibitemShut {NoStop}%
\bibitem [{\citenamefont {Schuld}\ \emph
  {et~al.}(2015{\natexlab{a}})\citenamefont {Schuld}, \citenamefont
  {Sinayskiy},\ and\ \citenamefont {Petruccione}}]{schuld2015simulating}%
  \BibitemOpen
  \bibfield  {author} {\bibinfo {author} {\bibfnamefont {M.}~\bibnamefont
  {Schuld}}, \bibinfo {author} {\bibfnamefont {I.}~\bibnamefont {Sinayskiy}},\
  and\ \bibinfo {author} {\bibfnamefont {F.}~\bibnamefont {Petruccione}},\
  }\bibfield  {title} {\bibinfo {title} {Simulating a perceptron on a quantum
  computer},\ }\href@noop {} {\bibfield  {journal} {\bibinfo  {journal}
  {Physics Letters A}\ }\textbf {\bibinfo {volume} {379}},\ \bibinfo {pages}
  {660} (\bibinfo {year} {2015}{\natexlab{a}})}\BibitemShut {NoStop}%
\bibitem [{\citenamefont {Espinosa-Ortega}\ and\ \citenamefont
  {Liew}(2015)}]{espinosa15}%
  \BibitemOpen
  \bibfield  {author} {\bibinfo {author} {\bibfnamefont {T.}~\bibnamefont
  {Espinosa-Ortega}}\ and\ \bibinfo {author} {\bibfnamefont {T.}~\bibnamefont
  {Liew}},\ }\bibfield  {title} {\bibinfo {title} {Perceptrons with {H}ebbian
  learning based on wave ensembles in spatially patterned potentials},\
  }\href@noop {} {\bibfield  {journal} {\bibinfo  {journal} {Physical Review
  Letters}\ }\textbf {\bibinfo {volume} {114}},\ \bibinfo {pages} {118101}
  (\bibinfo {year} {2015})}\BibitemShut {NoStop}%
\bibitem [{\citenamefont {Benatti}\ \emph {et~al.}(2019)\citenamefont
  {Benatti}, \citenamefont {Mancini},\ and\ \citenamefont
  {Mangini}}]{benatti2019continuous}%
  \BibitemOpen
  \bibfield  {author} {\bibinfo {author} {\bibfnamefont {F.}~\bibnamefont
  {Benatti}}, \bibinfo {author} {\bibfnamefont {S.}~\bibnamefont {Mancini}},\
  and\ \bibinfo {author} {\bibfnamefont {S.}~\bibnamefont {Mangini}},\
  }\bibfield  {title} {\bibinfo {title} {Continuous variable quantum
  perceptron},\ }\href@noop {} {\bibfield  {journal} {\bibinfo  {journal}
  {International Journal of Quantum Information}\ }\textbf {\bibinfo {volume}
  {17}},\ \bibinfo {pages} {1941009} (\bibinfo {year} {2019})}\BibitemShut
  {NoStop}%
\bibitem [{\citenamefont {Bondesan}\ and\ \citenamefont
  {Welling}(2020)}]{bondesan2020quantum}%
  \BibitemOpen
  \bibfield  {author} {\bibinfo {author} {\bibfnamefont {R.}~\bibnamefont
  {Bondesan}}\ and\ \bibinfo {author} {\bibfnamefont {M.}~\bibnamefont
  {Welling}},\ }\bibfield  {title} {\bibinfo {title} {Quantum deformed neural
  networks},\ }\href@noop {} {\bibfield  {journal} {\bibinfo  {journal} {arXiv
  preprint arXiv:2010.11189}\ } (\bibinfo {year} {2020})}\BibitemShut {NoStop}%
\bibitem [{\citenamefont {Chollet}(2019)}]{chollet2019measure}%
  \BibitemOpen
  \bibfield  {author} {\bibinfo {author} {\bibfnamefont {F.}~\bibnamefont
  {Chollet}},\ }\bibfield  {title} {\bibinfo {title} {On the measure of
  intelligence},\ }\href@noop {} {\bibfield  {journal} {\bibinfo  {journal}
  {arXiv preprint arXiv:1911.01547}\ } (\bibinfo {year} {2019})}\BibitemShut
  {NoStop}%
\bibitem [{\citenamefont {Bishop}\ and\ \citenamefont
  {Nasrabadi}(2006)}]{bishop2006pattern}%
  \BibitemOpen
  \bibfield  {author} {\bibinfo {author} {\bibfnamefont {C.~M.}\ \bibnamefont
  {Bishop}}\ and\ \bibinfo {author} {\bibfnamefont {N.~M.}\ \bibnamefont
  {Nasrabadi}},\ }\href@noop {} {\emph {\bibinfo {title} {Pattern recognition
  and machine learning}}},\ Vol.~\bibinfo {volume} {4}\ (\bibinfo  {publisher}
  {Springer},\ \bibinfo {year} {2006})\BibitemShut {NoStop}%
\bibitem [{\citenamefont {James}\ \emph {et~al.}(2013)\citenamefont {James},
  \citenamefont {Witten}, \citenamefont {Hastie},\ and\ \citenamefont
  {Tibshirani}}]{james2013introduction}%
  \BibitemOpen
  \bibfield  {author} {\bibinfo {author} {\bibfnamefont {G.}~\bibnamefont
  {James}}, \bibinfo {author} {\bibfnamefont {D.}~\bibnamefont {Witten}},
  \bibinfo {author} {\bibfnamefont {T.}~\bibnamefont {Hastie}},\ and\ \bibinfo
  {author} {\bibfnamefont {R.}~\bibnamefont {Tibshirani}},\ }\href@noop {}
  {\emph {\bibinfo {title} {An introduction to statistical learning}}},\ Vol.\
  \bibinfo {volume} {112}\ (\bibinfo  {publisher} {Springer},\ \bibinfo {year}
  {2013})\BibitemShut {NoStop}%
\bibitem [{\citenamefont {Goodfellow}\ \emph {et~al.}(2016)\citenamefont
  {Goodfellow}, \citenamefont {Bengio},\ and\ \citenamefont
  {Courville}}]{goodfellow2016deep}%
  \BibitemOpen
  \bibfield  {author} {\bibinfo {author} {\bibfnamefont {I.}~\bibnamefont
  {Goodfellow}}, \bibinfo {author} {\bibfnamefont {Y.}~\bibnamefont {Bengio}},\
  and\ \bibinfo {author} {\bibfnamefont {A.}~\bibnamefont {Courville}},\
  }\href@noop {} {\emph {\bibinfo {title} {Deep learning}}}\ (\bibinfo
  {publisher} {MIT press},\ \bibinfo {year} {2016})\BibitemShut {NoStop}%
\bibitem [{\citenamefont {Servedio}\ and\ \citenamefont
  {Gortler}(2004)}]{servedio04}%
  \BibitemOpen
  \bibfield  {author} {\bibinfo {author} {\bibfnamefont {R.~A.}\ \bibnamefont
  {Servedio}}\ and\ \bibinfo {author} {\bibfnamefont {S.~J.}\ \bibnamefont
  {Gortler}},\ }\bibfield  {title} {\bibinfo {title} {Equivalences and
  separations between quantum and classical learnability},\ }\href@noop {}
  {\bibfield  {journal} {\bibinfo  {journal} {SIAM Journal on Computing}\
  }\textbf {\bibinfo {volume} {33}},\ \bibinfo {pages} {1067} (\bibinfo {year}
  {2004})}\BibitemShut {NoStop}%
\bibitem [{\citenamefont {Zhang}\ \emph {et~al.}(2021)\citenamefont {Zhang},
  \citenamefont {Bengio}, \citenamefont {Hardt}, \citenamefont {Recht},\ and\
  \citenamefont {Vinyals}}]{zhang2021understanding}%
  \BibitemOpen
  \bibfield  {author} {\bibinfo {author} {\bibfnamefont {C.}~\bibnamefont
  {Zhang}}, \bibinfo {author} {\bibfnamefont {S.}~\bibnamefont {Bengio}},
  \bibinfo {author} {\bibfnamefont {M.}~\bibnamefont {Hardt}}, \bibinfo
  {author} {\bibfnamefont {B.}~\bibnamefont {Recht}},\ and\ \bibinfo {author}
  {\bibfnamefont {O.}~\bibnamefont {Vinyals}},\ }\bibfield  {title} {\bibinfo
  {title} {Understanding deep learning (still) requires rethinking
  generalization},\ }\href@noop {} {\bibfield  {journal} {\bibinfo  {journal}
  {Communications of the ACM}\ }\textbf {\bibinfo {volume} {64}},\ \bibinfo
  {pages} {107} (\bibinfo {year} {2021})}\BibitemShut {NoStop}%
\bibitem [{\citenamefont {Bartlett}\ \emph {et~al.}(2021)\citenamefont
  {Bartlett}, \citenamefont {Montanari},\ and\ \citenamefont
  {Rakhlin}}]{bartlett2021deep}%
  \BibitemOpen
  \bibfield  {author} {\bibinfo {author} {\bibfnamefont {P.~L.}\ \bibnamefont
  {Bartlett}}, \bibinfo {author} {\bibfnamefont {A.}~\bibnamefont
  {Montanari}},\ and\ \bibinfo {author} {\bibfnamefont {A.}~\bibnamefont
  {Rakhlin}},\ }\bibfield  {title} {\bibinfo {title} {Deep learning: a
  statistical viewpoint},\ }\href@noop {} {\bibfield  {journal} {\bibinfo
  {journal} {arXiv preprint arXiv:2103.09177}\ } (\bibinfo {year}
  {2021})}\BibitemShut {NoStop}%
\bibitem [{\citenamefont {Soudry}\ \emph {et~al.}(2018)\citenamefont {Soudry},
  \citenamefont {Hoffer}, \citenamefont {Nacson}, \citenamefont {Gunasekar},\
  and\ \citenamefont {Srebro}}]{soudry2018implicit}%
  \BibitemOpen
  \bibfield  {author} {\bibinfo {author} {\bibfnamefont {D.}~\bibnamefont
  {Soudry}}, \bibinfo {author} {\bibfnamefont {E.}~\bibnamefont {Hoffer}},
  \bibinfo {author} {\bibfnamefont {M.~S.}\ \bibnamefont {Nacson}}, \bibinfo
  {author} {\bibfnamefont {S.}~\bibnamefont {Gunasekar}},\ and\ \bibinfo
  {author} {\bibfnamefont {N.}~\bibnamefont {Srebro}},\ }\bibfield  {title}
  {\bibinfo {title} {The implicit bias of gradient descent on separable data},\
  }\href@noop {} {\bibfield  {journal} {\bibinfo  {journal} {The Journal of
  Machine Learning Research}\ }\textbf {\bibinfo {volume} {19}},\ \bibinfo
  {pages} {2822} (\bibinfo {year} {2018})}\BibitemShut {NoStop}%
\bibitem [{\citenamefont {Arora}\ \emph {et~al.}(2019)\citenamefont {Arora},
  \citenamefont {Cohen}, \citenamefont {Hu},\ and\ \citenamefont
  {Luo}}]{arora2019implicit}%
  \BibitemOpen
  \bibfield  {author} {\bibinfo {author} {\bibfnamefont {S.}~\bibnamefont
  {Arora}}, \bibinfo {author} {\bibfnamefont {N.}~\bibnamefont {Cohen}},
  \bibinfo {author} {\bibfnamefont {W.}~\bibnamefont {Hu}},\ and\ \bibinfo
  {author} {\bibfnamefont {Y.}~\bibnamefont {Luo}},\ }\bibfield  {title}
  {\bibinfo {title} {Implicit regularization in deep matrix factorization},\
  }\href@noop {} {\bibfield  {journal} {\bibinfo  {journal} {Advances in Neural
  Information Processing Systems}\ }\textbf {\bibinfo {volume} {32}},\ \bibinfo
  {pages} {7413} (\bibinfo {year} {2019})}\BibitemShut {NoStop}%
\bibitem [{\citenamefont {Kak}(1995)}]{kak95}%
  \BibitemOpen
  \bibfield  {author} {\bibinfo {author} {\bibfnamefont {S.~C.}\ \bibnamefont
  {Kak}},\ }\bibfield  {title} {\bibinfo {title} {Quantum neural computing},\
  }\href@noop {} {\bibfield  {journal} {\bibinfo  {journal} {Advances in
  Imaging and Electron Physics}\ }\textbf {\bibinfo {volume} {94}},\ \bibinfo
  {pages} {259} (\bibinfo {year} {1995})}\BibitemShut {NoStop}%
\bibitem [{\citenamefont {Bonnell}\ and\ \citenamefont
  {Papini}(1997)}]{bonnell97}%
  \BibitemOpen
  \bibfield  {author} {\bibinfo {author} {\bibfnamefont {G.}~\bibnamefont
  {Bonnell}}\ and\ \bibinfo {author} {\bibfnamefont {G.}~\bibnamefont
  {Papini}},\ }\bibfield  {title} {\bibinfo {title} {Quantum neural network},\
  }\href@noop {} {\bibfield  {journal} {\bibinfo  {journal} {International
  Journal of Theoretical Physics}\ }\textbf {\bibinfo {volume} {36}},\ \bibinfo
  {pages} {2855} (\bibinfo {year} {1997})}\BibitemShut {NoStop}%
\bibitem [{\citenamefont {Ventura}\ and\ \citenamefont
  {Martinez}(2000)}]{ventura00}%
  \BibitemOpen
  \bibfield  {author} {\bibinfo {author} {\bibfnamefont {D.}~\bibnamefont
  {Ventura}}\ and\ \bibinfo {author} {\bibfnamefont {T.}~\bibnamefont
  {Martinez}},\ }\bibfield  {title} {\bibinfo {title} {Quantum associative
  memory},\ }\href@noop {} {\bibfield  {journal} {\bibinfo  {journal}
  {Information Sciences}\ }\textbf {\bibinfo {volume} {124}},\ \bibinfo {pages}
  {273} (\bibinfo {year} {2000})}\BibitemShut {NoStop}%
\bibitem [{\citenamefont {Trugenberger}(2002)}]{trugenberger02}%
  \BibitemOpen
  \bibfield  {author} {\bibinfo {author} {\bibfnamefont {C.~A.}\ \bibnamefont
  {Trugenberger}},\ }\bibfield  {title} {\bibinfo {title} {Quantum pattern
  recognition},\ }\href@noop {} {\bibfield  {journal} {\bibinfo  {journal}
  {Quantum Information Processing}\ }\textbf {\bibinfo {volume} {1}},\ \bibinfo
  {pages} {471} (\bibinfo {year} {2002})}\BibitemShut {NoStop}%
\bibitem [{\citenamefont {Sch{\"u}tzhold}(2003)}]{schutzhold03}%
  \BibitemOpen
  \bibfield  {author} {\bibinfo {author} {\bibfnamefont {R.}~\bibnamefont
  {Sch{\"u}tzhold}},\ }\bibfield  {title} {\bibinfo {title} {Pattern
  recognition on a quantum computer},\ }\href@noop {} {\bibfield  {journal}
  {\bibinfo  {journal} {Physical Review A}\ }\textbf {\bibinfo {volume} {67}},\
  \bibinfo {pages} {062311} (\bibinfo {year} {2003})}\BibitemShut {NoStop}%
\bibitem [{\citenamefont {Bonner}\ and\ \citenamefont
  {Freivalds}(2003)}]{bonner03}%
  \BibitemOpen
  \bibfield  {author} {\bibinfo {author} {\bibfnamefont {R.}~\bibnamefont
  {Bonner}}\ and\ \bibinfo {author} {\bibfnamefont {R.}~\bibnamefont
  {Freivalds}},\ }\bibfield  {title} {\bibinfo {title} {A survey of quantum
  learning},\ }in\ \href@noop {} {\emph {\bibinfo {booktitle} {Quantum
  Computation and Learning}}}\ (\bibinfo {year} {2003})\ p.\ \bibinfo {pages}
  {106}\BibitemShut {NoStop}%
\bibitem [{\citenamefont {A{\"\i}meur}\ \emph {et~al.}(2006)\citenamefont
  {A{\"\i}meur}, \citenamefont {Brassard},\ and\ \citenamefont
  {Gambs}}]{aimeur06}%
  \BibitemOpen
  \bibfield  {author} {\bibinfo {author} {\bibfnamefont {E.}~\bibnamefont
  {A{\"\i}meur}}, \bibinfo {author} {\bibfnamefont {G.}~\bibnamefont
  {Brassard}},\ and\ \bibinfo {author} {\bibfnamefont {S.}~\bibnamefont
  {Gambs}},\ }\bibfield  {title} {\bibinfo {title} {Machine learning in a
  quantum world},\ }in\ \href@noop {} {\emph {\bibinfo {booktitle} {Advances in
  Artificial Intelligence}}}\ (\bibinfo  {publisher} {Springer},\ \bibinfo
  {year} {2006})\ pp.\ \bibinfo {pages} {431--442}\BibitemShut {NoStop}%
\bibitem [{\citenamefont {Bshouty}\ and\ \citenamefont
  {Jackson}(1998)}]{bshouty1998learning}%
  \BibitemOpen
  \bibfield  {author} {\bibinfo {author} {\bibfnamefont {N.~H.}\ \bibnamefont
  {Bshouty}}\ and\ \bibinfo {author} {\bibfnamefont {J.~C.}\ \bibnamefont
  {Jackson}},\ }\bibfield  {title} {\bibinfo {title} {Learning dnf over the
  uniform distribution using a quantum example oracle},\ }\href@noop {}
  {\bibfield  {journal} {\bibinfo  {journal} {SIAM Journal on Computing}\
  }\textbf {\bibinfo {volume} {28}},\ \bibinfo {pages} {1136} (\bibinfo {year}
  {1998})}\BibitemShut {NoStop}%
\bibitem [{\citenamefont {Schuld}\ \emph
  {et~al.}(2015{\natexlab{b}})\citenamefont {Schuld}, \citenamefont
  {Sinayskiy},\ and\ \citenamefont {Petruccione}}]{schuld15qml}%
  \BibitemOpen
  \bibfield  {author} {\bibinfo {author} {\bibfnamefont {M.}~\bibnamefont
  {Schuld}}, \bibinfo {author} {\bibfnamefont {I.}~\bibnamefont {Sinayskiy}},\
  and\ \bibinfo {author} {\bibfnamefont {F.}~\bibnamefont {Petruccione}},\
  }\bibfield  {title} {\bibinfo {title} {Introduction to quantum machine
  learning},\ }\href@noop {} {\bibfield  {journal} {\bibinfo  {journal}
  {Contemporary Physics}\ }\textbf {\bibinfo {volume} {56}},\ \bibinfo {pages}
  {172} (\bibinfo {year} {2015}{\natexlab{b}})}\BibitemShut {NoStop}%
\bibitem [{\citenamefont {Wiebe}\ \emph {et~al.}(2012)\citenamefont {Wiebe},
  \citenamefont {Braun},\ and\ \citenamefont {Lloyd}}]{wiebe12}%
  \BibitemOpen
  \bibfield  {author} {\bibinfo {author} {\bibfnamefont {N.}~\bibnamefont
  {Wiebe}}, \bibinfo {author} {\bibfnamefont {D.}~\bibnamefont {Braun}},\ and\
  \bibinfo {author} {\bibfnamefont {S.}~\bibnamefont {Lloyd}},\ }\bibfield
  {title} {\bibinfo {title} {Quantum algorithm for data fitting},\ }\href@noop
  {} {\bibfield  {journal} {\bibinfo  {journal} {Physical Review Letters}\
  }\textbf {\bibinfo {volume} {109}},\ \bibinfo {pages} {050505} (\bibinfo
  {year} {2012})}\BibitemShut {NoStop}%
\bibitem [{\citenamefont {Rebentrost}\ \emph {et~al.}(2014)\citenamefont
  {Rebentrost}, \citenamefont {Mohseni},\ and\ \citenamefont
  {Lloyd}}]{rebentrost14}%
  \BibitemOpen
  \bibfield  {author} {\bibinfo {author} {\bibfnamefont {P.}~\bibnamefont
  {Rebentrost}}, \bibinfo {author} {\bibfnamefont {M.}~\bibnamefont
  {Mohseni}},\ and\ \bibinfo {author} {\bibfnamefont {S.}~\bibnamefont
  {Lloyd}},\ }\bibfield  {title} {\bibinfo {title} {Quantum support vector
  machine for big data classification},\ }\href@noop {} {\bibfield  {journal}
  {\bibinfo  {journal} {Physcial Review Letters}\ }\textbf {\bibinfo {volume}
  {113}},\ \bibinfo {pages} {130503} (\bibinfo {year} {2014})}\BibitemShut
  {NoStop}%
\bibitem [{\citenamefont {Zlokapa}\ \emph {et~al.}(2021)\citenamefont
  {Zlokapa}, \citenamefont {Neven},\ and\ \citenamefont
  {Lloyd}}]{zlokapa2021quantum}%
  \BibitemOpen
  \bibfield  {author} {\bibinfo {author} {\bibfnamefont {A.}~\bibnamefont
  {Zlokapa}}, \bibinfo {author} {\bibfnamefont {H.}~\bibnamefont {Neven}},\
  and\ \bibinfo {author} {\bibfnamefont {S.}~\bibnamefont {Lloyd}},\ }\bibfield
   {title} {\bibinfo {title} {A quantum algorithm for training wide and deep
  classical neural networks},\ }\href@noop {} {\bibfield  {journal} {\bibinfo
  {journal} {arXiv preprint arXiv:2107.09200}\ } (\bibinfo {year}
  {2021})}\BibitemShut {NoStop}%
\bibitem [{\citenamefont {Wiebe}\ \emph {et~al.}(2014)\citenamefont {Wiebe},
  \citenamefont {Kapoor},\ and\ \citenamefont {Svore}}]{wiebe14b}%
  \BibitemOpen
  \bibfield  {author} {\bibinfo {author} {\bibfnamefont {N.}~\bibnamefont
  {Wiebe}}, \bibinfo {author} {\bibfnamefont {A.}~\bibnamefont {Kapoor}},\ and\
  \bibinfo {author} {\bibfnamefont {K.~M.}\ \bibnamefont {Svore}},\ }\href@noop
  {} {\bibinfo {title} {Quantum deep learning}} (\bibinfo {year} {2014}),\
  \bibinfo {note} {arXiv: 1412.3489v1}\BibitemShut {NoStop}%
\bibitem [{\citenamefont {Denil}\ and\ \citenamefont
  {De~Freitas}(2011)}]{denil11}%
  \BibitemOpen
  \bibfield  {author} {\bibinfo {author} {\bibfnamefont {M.}~\bibnamefont
  {Denil}}\ and\ \bibinfo {author} {\bibfnamefont {N.}~\bibnamefont
  {De~Freitas}},\ }\bibfield  {title} {\bibinfo {title} {Toward the
  implementation of a quantum {RBM}},\ }in\ \href@noop {} {\emph {\bibinfo
  {booktitle} {NIPS 2011 Deep Learning and Unsupervised Feature Learning
  Workshop}}}\ (\bibinfo {year} {2011})\BibitemShut {NoStop}%
\bibitem [{\citenamefont {Kerenidis}\ and\ \citenamefont
  {Prakash}(2022)}]{kerenidis2022quantum}%
  \BibitemOpen
  \bibfield  {author} {\bibinfo {author} {\bibfnamefont {I.}~\bibnamefont
  {Kerenidis}}\ and\ \bibinfo {author} {\bibfnamefont {A.}~\bibnamefont
  {Prakash}},\ }\bibfield  {title} {\bibinfo {title} {Quantum machine learning
  with subspace states},\ }\href@noop {} {\bibfield  {journal} {\bibinfo
  {journal} {arXiv preprint arXiv:2202.00054}\ } (\bibinfo {year}
  {2022})}\BibitemShut {NoStop}%
\bibitem [{\citenamefont {A{\"\i}meur}\ \emph {et~al.}(2013)\citenamefont
  {A{\"\i}meur}, \citenamefont {Brassard},\ and\ \citenamefont
  {Gambs}}]{aimeur13}%
  \BibitemOpen
  \bibfield  {author} {\bibinfo {author} {\bibfnamefont {E.}~\bibnamefont
  {A{\"\i}meur}}, \bibinfo {author} {\bibfnamefont {G.}~\bibnamefont
  {Brassard}},\ and\ \bibinfo {author} {\bibfnamefont {S.}~\bibnamefont
  {Gambs}},\ }\bibfield  {title} {\bibinfo {title} {Quantum speed-up for
  unsupervised learning},\ }\href@noop {} {\bibfield  {journal} {\bibinfo
  {journal} {Machine Learning}\ }\textbf {\bibinfo {volume} {90}},\ \bibinfo
  {pages} {261} (\bibinfo {year} {2013})}\BibitemShut {NoStop}%
\bibitem [{\citenamefont {Farhi}\ and\ \citenamefont {Neven}(2018)}]{farhi18}%
  \BibitemOpen
  \bibfield  {author} {\bibinfo {author} {\bibfnamefont {E.}~\bibnamefont
  {Farhi}}\ and\ \bibinfo {author} {\bibfnamefont {H.}~\bibnamefont {Neven}},\
  }\bibfield  {title} {\bibinfo {title} {Classification with quantum neural
  networks on near term processors},\ }\href@noop {} {\bibfield  {journal}
  {\bibinfo  {journal} {arXiv preprint arXiv:1802.06002}\ } (\bibinfo {year}
  {2018})}\BibitemShut {NoStop}%
\bibitem [{\citenamefont {Schuld}\ \emph {et~al.}(2018)\citenamefont {Schuld},
  \citenamefont {Bocharov}, \citenamefont {Svore},\ and\ \citenamefont
  {Wiebe}}]{schuld18cc}%
  \BibitemOpen
  \bibfield  {author} {\bibinfo {author} {\bibfnamefont {M.}~\bibnamefont
  {Schuld}}, \bibinfo {author} {\bibfnamefont {A.}~\bibnamefont {Bocharov}},
  \bibinfo {author} {\bibfnamefont {K.}~\bibnamefont {Svore}},\ and\ \bibinfo
  {author} {\bibfnamefont {N.}~\bibnamefont {Wiebe}},\ }\bibfield  {title}
  {\bibinfo {title} {Circuit-centric quantum classifiers},\ }\href@noop {}
  {\bibfield  {journal} {\bibinfo  {journal} {arXiv preprint arXiv:1804.00633}\
  } (\bibinfo {year} {2018})}\BibitemShut {NoStop}%
\bibitem [{\citenamefont {Liu}\ and\ \citenamefont
  {Wang}(2018)}]{liu2018differentiable}%
  \BibitemOpen
  \bibfield  {author} {\bibinfo {author} {\bibfnamefont {J.-G.}\ \bibnamefont
  {Liu}}\ and\ \bibinfo {author} {\bibfnamefont {L.}~\bibnamefont {Wang}},\
  }\bibfield  {title} {\bibinfo {title} {Differentiable learning of quantum
  circuit born machines},\ }\href@noop {} {\bibfield  {journal} {\bibinfo
  {journal} {Physical Review A}\ }\textbf {\bibinfo {volume} {98}},\ \bibinfo
  {pages} {062324} (\bibinfo {year} {2018})}\BibitemShut {NoStop}%
\bibitem [{\citenamefont {Benedetti}\ \emph {et~al.}(2019)\citenamefont
  {Benedetti}, \citenamefont {Lloyd}, \citenamefont {Sack},\ and\ \citenamefont
  {Fiorentini}}]{benedetti2019parameterized}%
  \BibitemOpen
  \bibfield  {author} {\bibinfo {author} {\bibfnamefont {M.}~\bibnamefont
  {Benedetti}}, \bibinfo {author} {\bibfnamefont {E.}~\bibnamefont {Lloyd}},
  \bibinfo {author} {\bibfnamefont {S.}~\bibnamefont {Sack}},\ and\ \bibinfo
  {author} {\bibfnamefont {M.}~\bibnamefont {Fiorentini}},\ }\bibfield  {title}
  {\bibinfo {title} {Parameterized quantum circuits as machine learning
  models},\ }\href@noop {} {\bibfield  {journal} {\bibinfo  {journal} {Quantum
  Science and Technology}\ }\textbf {\bibinfo {volume} {4}},\ \bibinfo {pages}
  {043001} (\bibinfo {year} {2019})}\BibitemShut {NoStop}%
\bibitem [{\citenamefont {Cong}\ \emph {et~al.}(2019)\citenamefont {Cong},
  \citenamefont {Choi},\ and\ \citenamefont {Lukin}}]{cong2019quantum}%
  \BibitemOpen
  \bibfield  {author} {\bibinfo {author} {\bibfnamefont {I.}~\bibnamefont
  {Cong}}, \bibinfo {author} {\bibfnamefont {S.}~\bibnamefont {Choi}},\ and\
  \bibinfo {author} {\bibfnamefont {M.~D.}\ \bibnamefont {Lukin}},\ }\bibfield
  {title} {\bibinfo {title} {Quantum convolutional neural networks},\
  }\href@noop {} {\bibfield  {journal} {\bibinfo  {journal} {Nature Physics}\
  }\textbf {\bibinfo {volume} {15}},\ \bibinfo {pages} {1273} (\bibinfo {year}
  {2019})}\BibitemShut {NoStop}%
\bibitem [{\citenamefont {P{\'e}rez-Salinas}\ \emph {et~al.}(2020)\citenamefont
  {P{\'e}rez-Salinas}, \citenamefont {Cervera-Lierta}, \citenamefont
  {Gil-Fuster},\ and\ \citenamefont {Latorre}}]{perez2020data}%
  \BibitemOpen
  \bibfield  {author} {\bibinfo {author} {\bibfnamefont {A.}~\bibnamefont
  {P{\'e}rez-Salinas}}, \bibinfo {author} {\bibfnamefont {A.}~\bibnamefont
  {Cervera-Lierta}}, \bibinfo {author} {\bibfnamefont {E.}~\bibnamefont
  {Gil-Fuster}},\ and\ \bibinfo {author} {\bibfnamefont {J.~I.}\ \bibnamefont
  {Latorre}},\ }\bibfield  {title} {\bibinfo {title} {Data re-uploading for a
  universal quantum classifier},\ }\href@noop {} {\bibfield  {journal}
  {\bibinfo  {journal} {Quantum}\ }\textbf {\bibinfo {volume} {4}},\ \bibinfo
  {pages} {226} (\bibinfo {year} {2020})}\BibitemShut {NoStop}%
\bibitem [{\citenamefont {Mitarai}\ \emph {et~al.}(2018)\citenamefont
  {Mitarai}, \citenamefont {Negoro}, \citenamefont {Kitagawa},\ and\
  \citenamefont {Fujii}}]{mitarai2018quantum}%
  \BibitemOpen
  \bibfield  {author} {\bibinfo {author} {\bibfnamefont {K.}~\bibnamefont
  {Mitarai}}, \bibinfo {author} {\bibfnamefont {M.}~\bibnamefont {Negoro}},
  \bibinfo {author} {\bibfnamefont {M.}~\bibnamefont {Kitagawa}},\ and\
  \bibinfo {author} {\bibfnamefont {K.}~\bibnamefont {Fujii}},\ }\bibfield
  {title} {\bibinfo {title} {Quantum circuit learning},\ }\href@noop {}
  {\bibfield  {journal} {\bibinfo  {journal} {Physical Review A}\ }\textbf
  {\bibinfo {volume} {98}},\ \bibinfo {pages} {032309} (\bibinfo {year}
  {2018})}\BibitemShut {NoStop}%
\bibitem [{\citenamefont {Schuld}\ \emph {et~al.}(2019)\citenamefont {Schuld},
  \citenamefont {Bergholm}, \citenamefont {Gogolin}, \citenamefont {Izaac},\
  and\ \citenamefont {Killoran}}]{schuld2019evaluating}%
  \BibitemOpen
  \bibfield  {author} {\bibinfo {author} {\bibfnamefont {M.}~\bibnamefont
  {Schuld}}, \bibinfo {author} {\bibfnamefont {V.}~\bibnamefont {Bergholm}},
  \bibinfo {author} {\bibfnamefont {C.}~\bibnamefont {Gogolin}}, \bibinfo
  {author} {\bibfnamefont {J.}~\bibnamefont {Izaac}},\ and\ \bibinfo {author}
  {\bibfnamefont {N.}~\bibnamefont {Killoran}},\ }\bibfield  {title} {\bibinfo
  {title} {Evaluating analytic gradients on quantum hardware},\ }\href@noop {}
  {\bibfield  {journal} {\bibinfo  {journal} {Physical Review A}\ }\textbf
  {\bibinfo {volume} {99}},\ \bibinfo {pages} {032331} (\bibinfo {year}
  {2019})}\BibitemShut {NoStop}%
\bibitem [{\citenamefont {McClean}\ \emph {et~al.}(2018)\citenamefont
  {McClean}, \citenamefont {Boixo}, \citenamefont {Smelyanskiy}, \citenamefont
  {Babbush},\ and\ \citenamefont {Neven}}]{mcclean2018barren}%
  \BibitemOpen
  \bibfield  {author} {\bibinfo {author} {\bibfnamefont {J.~R.}\ \bibnamefont
  {McClean}}, \bibinfo {author} {\bibfnamefont {S.}~\bibnamefont {Boixo}},
  \bibinfo {author} {\bibfnamefont {V.~N.}\ \bibnamefont {Smelyanskiy}},
  \bibinfo {author} {\bibfnamefont {R.}~\bibnamefont {Babbush}},\ and\ \bibinfo
  {author} {\bibfnamefont {H.}~\bibnamefont {Neven}},\ }\bibfield  {title}
  {\bibinfo {title} {Barren plateaus in quantum neural network training
  landscapes},\ }\href@noop {} {\bibfield  {journal} {\bibinfo  {journal}
  {Nature communications}\ }\textbf {\bibinfo {volume} {9}},\ \bibinfo {pages}
  {1} (\bibinfo {year} {2018})}\BibitemShut {NoStop}%
\bibitem [{\citenamefont {Holmes}\ \emph {et~al.}(2021)\citenamefont {Holmes},
  \citenamefont {Sharma}, \citenamefont {Cerezo},\ and\ \citenamefont
  {Coles}}]{holmes2021connecting}%
  \BibitemOpen
  \bibfield  {author} {\bibinfo {author} {\bibfnamefont {Z.}~\bibnamefont
  {Holmes}}, \bibinfo {author} {\bibfnamefont {K.}~\bibnamefont {Sharma}},
  \bibinfo {author} {\bibfnamefont {M.}~\bibnamefont {Cerezo}},\ and\ \bibinfo
  {author} {\bibfnamefont {P.~J.}\ \bibnamefont {Coles}},\ }\bibfield  {title}
  {\bibinfo {title} {Connecting ansatz expressibility to gradient magnitudes
  and barren plateaus},\ }\href@noop {} {\bibfield  {journal} {\bibinfo
  {journal} {arXiv preprint arXiv:2101.02138}\ } (\bibinfo {year}
  {2021})}\BibitemShut {NoStop}%
\bibitem [{\citenamefont {Wu}\ \emph {et~al.}(2021)\citenamefont {Wu},
  \citenamefont {Yao}, \citenamefont {Zhang},\ and\ \citenamefont
  {Zhai}}]{wu2021expressivity}%
  \BibitemOpen
  \bibfield  {author} {\bibinfo {author} {\bibfnamefont {Y.}~\bibnamefont
  {Wu}}, \bibinfo {author} {\bibfnamefont {J.}~\bibnamefont {Yao}}, \bibinfo
  {author} {\bibfnamefont {P.}~\bibnamefont {Zhang}},\ and\ \bibinfo {author}
  {\bibfnamefont {H.}~\bibnamefont {Zhai}},\ }\bibfield  {title} {\bibinfo
  {title} {Expressivity of quantum neural networks},\ }\href@noop {} {\bibfield
   {journal} {\bibinfo  {journal} {Physical Review Research}\ }\textbf
  {\bibinfo {volume} {3}},\ \bibinfo {pages} {L032049} (\bibinfo {year}
  {2021})}\BibitemShut {NoStop}%
\bibitem [{\citenamefont {Schuld}\ \emph {et~al.}(2021)\citenamefont {Schuld},
  \citenamefont {Sweke},\ and\ \citenamefont {Meyer}}]{schuld2021effect}%
  \BibitemOpen
  \bibfield  {author} {\bibinfo {author} {\bibfnamefont {M.}~\bibnamefont
  {Schuld}}, \bibinfo {author} {\bibfnamefont {R.}~\bibnamefont {Sweke}},\ and\
  \bibinfo {author} {\bibfnamefont {J.~J.}\ \bibnamefont {Meyer}},\ }\bibfield
  {title} {\bibinfo {title} {Effect of data encoding on the expressive power of
  variational quantum-machine-learning models},\ }\href@noop {} {\bibfield
  {journal} {\bibinfo  {journal} {Physical Review A}\ }\textbf {\bibinfo
  {volume} {103}},\ \bibinfo {pages} {032430} (\bibinfo {year}
  {2021})}\BibitemShut {NoStop}%
\bibitem [{\citenamefont {Huang}\ \emph
  {et~al.}(2021{\natexlab{b}})\citenamefont {Huang}, \citenamefont {Broughton},
  \citenamefont {Mohseni}, \citenamefont {Babbush}, \citenamefont {Boixo},
  \citenamefont {Neven},\ and\ \citenamefont {McClean}}]{huang2021power}%
  \BibitemOpen
  \bibfield  {author} {\bibinfo {author} {\bibfnamefont {H.-Y.}\ \bibnamefont
  {Huang}}, \bibinfo {author} {\bibfnamefont {M.}~\bibnamefont {Broughton}},
  \bibinfo {author} {\bibfnamefont {M.}~\bibnamefont {Mohseni}}, \bibinfo
  {author} {\bibfnamefont {R.}~\bibnamefont {Babbush}}, \bibinfo {author}
  {\bibfnamefont {S.}~\bibnamefont {Boixo}}, \bibinfo {author} {\bibfnamefont
  {H.}~\bibnamefont {Neven}},\ and\ \bibinfo {author} {\bibfnamefont {J.~R.}\
  \bibnamefont {McClean}},\ }\bibfield  {title} {\bibinfo {title} {Power of
  data in quantum machine learning},\ }\href@noop {} {\bibfield  {journal}
  {\bibinfo  {journal} {Nature communications}\ }\textbf {\bibinfo {volume}
  {12}},\ \bibinfo {pages} {1} (\bibinfo {year}
  {2021}{\natexlab{b}})}\BibitemShut {NoStop}%
\bibitem [{\citenamefont {Abbas}\ \emph {et~al.}(2021)\citenamefont {Abbas},
  \citenamefont {Sutter}, \citenamefont {Zoufal}, \citenamefont {Lucchi},
  \citenamefont {Figalli},\ and\ \citenamefont {Woerner}}]{abbas2021power}%
  \BibitemOpen
  \bibfield  {author} {\bibinfo {author} {\bibfnamefont {A.}~\bibnamefont
  {Abbas}}, \bibinfo {author} {\bibfnamefont {D.}~\bibnamefont {Sutter}},
  \bibinfo {author} {\bibfnamefont {C.}~\bibnamefont {Zoufal}}, \bibinfo
  {author} {\bibfnamefont {A.}~\bibnamefont {Lucchi}}, \bibinfo {author}
  {\bibfnamefont {A.}~\bibnamefont {Figalli}},\ and\ \bibinfo {author}
  {\bibfnamefont {S.}~\bibnamefont {Woerner}},\ }\bibfield  {title} {\bibinfo
  {title} {The power of quantum neural networks},\ }\href@noop {} {\bibfield
  {journal} {\bibinfo  {journal} {Nature Computational Science}\ }\textbf
  {\bibinfo {volume} {1}},\ \bibinfo {pages} {403} (\bibinfo {year}
  {2021})}\BibitemShut {NoStop}%
\bibitem [{\citenamefont {Caro}\ \emph {et~al.}(2021)\citenamefont {Caro},
  \citenamefont {Huang}, \citenamefont {Cerezo}, \citenamefont {Sharma},
  \citenamefont {Sornborger}, \citenamefont {Cincio},\ and\ \citenamefont
  {Coles}}]{caro2021generalization}%
  \BibitemOpen
  \bibfield  {author} {\bibinfo {author} {\bibfnamefont {M.~C.}\ \bibnamefont
  {Caro}}, \bibinfo {author} {\bibfnamefont {H.-Y.}\ \bibnamefont {Huang}},
  \bibinfo {author} {\bibfnamefont {M.}~\bibnamefont {Cerezo}}, \bibinfo
  {author} {\bibfnamefont {K.}~\bibnamefont {Sharma}}, \bibinfo {author}
  {\bibfnamefont {A.}~\bibnamefont {Sornborger}}, \bibinfo {author}
  {\bibfnamefont {L.}~\bibnamefont {Cincio}},\ and\ \bibinfo {author}
  {\bibfnamefont {P.~J.}\ \bibnamefont {Coles}},\ }\bibfield  {title} {\bibinfo
  {title} {Generalization in quantum machine learning from few training data},\
  }\href@noop {} {\bibfield  {journal} {\bibinfo  {journal} {arXiv preprint
  arXiv:2111.05292}\ } (\bibinfo {year} {2021})}\BibitemShut {NoStop}%
\bibitem [{\citenamefont {Banchi}\ \emph {et~al.}(2021)\citenamefont {Banchi},
  \citenamefont {Pereira},\ and\ \citenamefont
  {Pirandola}}]{banchi2021generalization}%
  \BibitemOpen
  \bibfield  {author} {\bibinfo {author} {\bibfnamefont {L.}~\bibnamefont
  {Banchi}}, \bibinfo {author} {\bibfnamefont {J.}~\bibnamefont {Pereira}},\
  and\ \bibinfo {author} {\bibfnamefont {S.}~\bibnamefont {Pirandola}},\
  }\bibfield  {title} {\bibinfo {title} {Generalization in quantum machine
  learning: a quantum information perspective},\ }\href@noop {} {\bibfield
  {journal} {\bibinfo  {journal} {arXiv preprint arXiv:2102.08991}\ } (\bibinfo
  {year} {2021})}\BibitemShut {NoStop}%
\bibitem [{\citenamefont {Arunachalam}\ and\ \citenamefont
  {de~Wolf}(2018)}]{arunachalam2018optimal}%
  \BibitemOpen
  \bibfield  {author} {\bibinfo {author} {\bibfnamefont {S.}~\bibnamefont
  {Arunachalam}}\ and\ \bibinfo {author} {\bibfnamefont {R.}~\bibnamefont
  {de~Wolf}},\ }\bibfield  {title} {\bibinfo {title} {Optimal quantum sample
  complexity of learning algorithms},\ }\href@noop {} {\bibfield  {journal}
  {\bibinfo  {journal} {The Journal of Machine Learning Research}\ }\textbf
  {\bibinfo {volume} {19}},\ \bibinfo {pages} {2879} (\bibinfo {year}
  {2018})}\BibitemShut {NoStop}%
\bibitem [{\citenamefont {Sent{\'\i}s}\ \emph {et~al.}(2012)\citenamefont
  {Sent{\'\i}s}, \citenamefont {Calsamiglia}, \citenamefont {Mu{\~n}oz-Tapia},\
  and\ \citenamefont {Bagan}}]{sentis12}%
  \BibitemOpen
  \bibfield  {author} {\bibinfo {author} {\bibfnamefont {G.}~\bibnamefont
  {Sent{\'\i}s}}, \bibinfo {author} {\bibfnamefont {J.}~\bibnamefont
  {Calsamiglia}}, \bibinfo {author} {\bibfnamefont {R.}~\bibnamefont
  {Mu{\~n}oz-Tapia}},\ and\ \bibinfo {author} {\bibfnamefont {E.}~\bibnamefont
  {Bagan}},\ }\bibfield  {title} {\bibinfo {title} {Quantum learning without
  quantum memory},\ }\href@noop {} {\bibfield  {journal} {\bibinfo  {journal}
  {Scientific Reports}\ }\textbf {\bibinfo {volume} {2}},\ \bibinfo {pages} {1}
  (\bibinfo {year} {2012})}\BibitemShut {NoStop}%
\bibitem [{\citenamefont {Monras}\ \emph {et~al.}(2017)\citenamefont {Monras},
  \citenamefont {Sent{\'\i}s},\ and\ \citenamefont
  {Wittek}}]{monras2017inductive}%
  \BibitemOpen
  \bibfield  {author} {\bibinfo {author} {\bibfnamefont {A.}~\bibnamefont
  {Monras}}, \bibinfo {author} {\bibfnamefont {G.}~\bibnamefont
  {Sent{\'\i}s}},\ and\ \bibinfo {author} {\bibfnamefont {P.}~\bibnamefont
  {Wittek}},\ }\bibfield  {title} {\bibinfo {title} {Inductive supervised
  quantum learning},\ }\href@noop {} {\bibfield  {journal} {\bibinfo  {journal}
  {Physical review letters}\ }\textbf {\bibinfo {volume} {118}},\ \bibinfo
  {pages} {190503} (\bibinfo {year} {2017})}\BibitemShut {NoStop}%
\bibitem [{\citenamefont {Paparo}\ \emph
  {et~al.}(2014{\natexlab{a}})\citenamefont {Paparo}, \citenamefont {Dunjko},
  \citenamefont {Makmal}, \citenamefont {Martin-Delgado},\ and\ \citenamefont
  {Briegel}}]{paparo14}%
  \BibitemOpen
  \bibfield  {author} {\bibinfo {author} {\bibfnamefont {G.~D.}\ \bibnamefont
  {Paparo}}, \bibinfo {author} {\bibfnamefont {V.}~\bibnamefont {Dunjko}},
  \bibinfo {author} {\bibfnamefont {A.}~\bibnamefont {Makmal}}, \bibinfo
  {author} {\bibfnamefont {M.~A.}\ \bibnamefont {Martin-Delgado}},\ and\
  \bibinfo {author} {\bibfnamefont {H.~J.}\ \bibnamefont {Briegel}},\
  }\bibfield  {title} {\bibinfo {title} {Quantum speedup for active learning
  agents},\ }\href@noop {} {\bibfield  {journal} {\bibinfo  {journal} {Physical
  Review X}\ }\textbf {\bibinfo {volume} {4}},\ \bibinfo {pages} {031002}
  (\bibinfo {year} {2014}{\natexlab{a}})}\BibitemShut {NoStop}%
\bibitem [{\citenamefont {Dunjko}\ \emph {et~al.}(2016)\citenamefont {Dunjko},
  \citenamefont {Taylor},\ and\ \citenamefont {Briegel}}]{dunjko16}%
  \BibitemOpen
  \bibfield  {author} {\bibinfo {author} {\bibfnamefont {V.}~\bibnamefont
  {Dunjko}}, \bibinfo {author} {\bibfnamefont {J.~M.}\ \bibnamefont {Taylor}},\
  and\ \bibinfo {author} {\bibfnamefont {H.~J.}\ \bibnamefont {Briegel}},\
  }\bibfield  {title} {\bibinfo {title} {Quantum-enhanced machine learning},\
  }\href@noop {} {\bibfield  {journal} {\bibinfo  {journal} {Physical Review
  Letters}\ }\textbf {\bibinfo {volume} {117}},\ \bibinfo {pages} {130501}
  (\bibinfo {year} {2016})}\BibitemShut {NoStop}%
\bibitem [{\citenamefont {Rotondo}\ \emph {et~al.}(2018)\citenamefont
  {Rotondo}, \citenamefont {Marcuzzi}, \citenamefont {Garrahan}, \citenamefont
  {Lesanovsky},\ and\ \citenamefont {M{\"u}ller}}]{rotondo2018open}%
  \BibitemOpen
  \bibfield  {author} {\bibinfo {author} {\bibfnamefont {P.}~\bibnamefont
  {Rotondo}}, \bibinfo {author} {\bibfnamefont {M.}~\bibnamefont {Marcuzzi}},
  \bibinfo {author} {\bibfnamefont {J.~P.}\ \bibnamefont {Garrahan}}, \bibinfo
  {author} {\bibfnamefont {I.}~\bibnamefont {Lesanovsky}},\ and\ \bibinfo
  {author} {\bibfnamefont {M.}~\bibnamefont {M{\"u}ller}},\ }\bibfield  {title}
  {\bibinfo {title} {Open quantum generalisation of hopfield neural networks},\
  }\href@noop {} {\bibfield  {journal} {\bibinfo  {journal} {Journal of Physics
  A: Mathematical and Theoretical}\ }\textbf {\bibinfo {volume} {51}},\
  \bibinfo {pages} {115301} (\bibinfo {year} {2018})}\BibitemShut {NoStop}%
\bibitem [{\citenamefont {Liu}\ \emph {et~al.}(2021{\natexlab{a}})\citenamefont
  {Liu}, \citenamefont {Arunachalam},\ and\ \citenamefont
  {Temme}}]{liu2021rigorous}%
  \BibitemOpen
  \bibfield  {author} {\bibinfo {author} {\bibfnamefont {Y.}~\bibnamefont
  {Liu}}, \bibinfo {author} {\bibfnamefont {S.}~\bibnamefont {Arunachalam}},\
  and\ \bibinfo {author} {\bibfnamefont {K.}~\bibnamefont {Temme}},\ }\bibfield
   {title} {\bibinfo {title} {A rigorous and robust quantum speed-up in
  supervised machine learning},\ }\href@noop {} {\bibfield  {journal} {\bibinfo
   {journal} {Nature Physics}\ ,\ \bibinfo {pages} {1}} (\bibinfo {year}
  {2021}{\natexlab{a}})}\BibitemShut {NoStop}%
\bibitem [{\citenamefont {Sweke}\ \emph {et~al.}(2021)\citenamefont {Sweke},
  \citenamefont {Seifert}, \citenamefont {Hangleiter},\ and\ \citenamefont
  {Eisert}}]{sweke2021quantum}%
  \BibitemOpen
  \bibfield  {author} {\bibinfo {author} {\bibfnamefont {R.}~\bibnamefont
  {Sweke}}, \bibinfo {author} {\bibfnamefont {J.-P.}\ \bibnamefont {Seifert}},
  \bibinfo {author} {\bibfnamefont {D.}~\bibnamefont {Hangleiter}},\ and\
  \bibinfo {author} {\bibfnamefont {J.}~\bibnamefont {Eisert}},\ }\bibfield
  {title} {\bibinfo {title} {On the quantum versus classical learnability of
  discrete distributions},\ }\href@noop {} {\bibfield  {journal} {\bibinfo
  {journal} {Quantum}\ }\textbf {\bibinfo {volume} {5}},\ \bibinfo {pages}
  {417} (\bibinfo {year} {2021})}\BibitemShut {NoStop}%
\bibitem [{\citenamefont {Du}\ \emph {et~al.}(2020)\citenamefont {Du},
  \citenamefont {Hsieh}, \citenamefont {Liu},\ and\ \citenamefont
  {Tao}}]{du2020expressive}%
  \BibitemOpen
  \bibfield  {author} {\bibinfo {author} {\bibfnamefont {Y.}~\bibnamefont
  {Du}}, \bibinfo {author} {\bibfnamefont {M.-H.}\ \bibnamefont {Hsieh}},
  \bibinfo {author} {\bibfnamefont {T.}~\bibnamefont {Liu}},\ and\ \bibinfo
  {author} {\bibfnamefont {D.}~\bibnamefont {Tao}},\ }\bibfield  {title}
  {\bibinfo {title} {Expressive power of parametrized quantum circuits},\
  }\href@noop {} {\bibfield  {journal} {\bibinfo  {journal} {Physical Review
  Research}\ }\textbf {\bibinfo {volume} {2}},\ \bibinfo {pages} {033125}
  (\bibinfo {year} {2020})}\BibitemShut {NoStop}%
\bibitem [{\citenamefont {Huang}\ \emph
  {et~al.}(2021{\natexlab{c}})\citenamefont {Huang}, \citenamefont {Broughton},
  \citenamefont {Cotler}, \citenamefont {Chen}, \citenamefont {Li},
  \citenamefont {Mohseni}, \citenamefont {Neven}, \citenamefont {Babbush},
  \citenamefont {Kueng}, \citenamefont {Preskill} \emph
  {et~al.}}]{huang2021quantum}%
  \BibitemOpen
  \bibfield  {author} {\bibinfo {author} {\bibfnamefont {H.-Y.}\ \bibnamefont
  {Huang}}, \bibinfo {author} {\bibfnamefont {M.}~\bibnamefont {Broughton}},
  \bibinfo {author} {\bibfnamefont {J.}~\bibnamefont {Cotler}}, \bibinfo
  {author} {\bibfnamefont {S.}~\bibnamefont {Chen}}, \bibinfo {author}
  {\bibfnamefont {J.}~\bibnamefont {Li}}, \bibinfo {author} {\bibfnamefont
  {M.}~\bibnamefont {Mohseni}}, \bibinfo {author} {\bibfnamefont
  {H.}~\bibnamefont {Neven}}, \bibinfo {author} {\bibfnamefont
  {R.}~\bibnamefont {Babbush}}, \bibinfo {author} {\bibfnamefont
  {R.}~\bibnamefont {Kueng}}, \bibinfo {author} {\bibfnamefont
  {J.}~\bibnamefont {Preskill}}, \emph {et~al.},\ }\bibfield  {title} {\bibinfo
  {title} {Quantum advantage in learning from experiments},\ }\href@noop {}
  {\bibfield  {journal} {\bibinfo  {journal} {arXiv preprint arXiv:2112.00778}\
  } (\bibinfo {year} {2021}{\natexlab{c}})}\BibitemShut {NoStop}%
\bibitem [{\citenamefont {Gili}\ \emph {et~al.}(2022)\citenamefont {Gili},
  \citenamefont {Mauri},\ and\ \citenamefont
  {Perdomo-Ortiz}}]{gili2022evaluating}%
  \BibitemOpen
  \bibfield  {author} {\bibinfo {author} {\bibfnamefont {K.}~\bibnamefont
  {Gili}}, \bibinfo {author} {\bibfnamefont {M.}~\bibnamefont {Mauri}},\ and\
  \bibinfo {author} {\bibfnamefont {A.}~\bibnamefont {Perdomo-Ortiz}},\
  }\bibfield  {title} {\bibinfo {title} {Evaluating generalization in classical
  and quantum generative models},\ }\href@noop {} {\bibfield  {journal}
  {\bibinfo  {journal} {arXiv preprint arXiv:2201.08770}\ } (\bibinfo {year}
  {2022})}\BibitemShut {NoStop}%
\bibitem [{\citenamefont {Mlinari{\'c}}\ \emph {et~al.}(2017)\citenamefont
  {Mlinari{\'c}}, \citenamefont {Horvat},\ and\ \citenamefont
  {{\v{S}}upak~Smol{\v{c}}i{\'c}}}]{mlinaric2017dealing}%
  \BibitemOpen
  \bibfield  {author} {\bibinfo {author} {\bibfnamefont {A.}~\bibnamefont
  {Mlinari{\'c}}}, \bibinfo {author} {\bibfnamefont {M.}~\bibnamefont
  {Horvat}},\ and\ \bibinfo {author} {\bibfnamefont {V.}~\bibnamefont
  {{\v{S}}upak~Smol{\v{c}}i{\'c}}},\ }\bibfield  {title} {\bibinfo {title}
  {Dealing with the positive publication bias: Why you should really publish
  your negative results},\ }\href@noop {} {\bibfield  {journal} {\bibinfo
  {journal} {Biochemia medica}\ }\textbf {\bibinfo {volume} {27}},\ \bibinfo
  {pages} {447} (\bibinfo {year} {2017})}\BibitemShut {NoStop}%
\bibitem [{\citenamefont {Alcazar}\ and\ \citenamefont
  {Perdomo-Ortiz}(2021)}]{alcazar2021enhancing}%
  \BibitemOpen
  \bibfield  {author} {\bibinfo {author} {\bibfnamefont {J.}~\bibnamefont
  {Alcazar}}\ and\ \bibinfo {author} {\bibfnamefont {A.}~\bibnamefont
  {Perdomo-Ortiz}},\ }\bibfield  {title} {\bibinfo {title} {Enhancing
  combinatorial optimization with quantum generative models},\ }\href@noop {}
  {\bibfield  {journal} {\bibinfo  {journal} {arXiv preprint arXiv:2101.06250}\
  } (\bibinfo {year} {2021})}\BibitemShut {NoStop}%
\bibitem [{\citenamefont {K{\"u}bler}\ \emph {et~al.}(2021)\citenamefont
  {K{\"u}bler}, \citenamefont {Buchholz},\ and\ \citenamefont
  {Sch{\"o}lkopf}}]{kubler2021inductive}%
  \BibitemOpen
  \bibfield  {author} {\bibinfo {author} {\bibfnamefont {J.}~\bibnamefont
  {K{\"u}bler}}, \bibinfo {author} {\bibfnamefont {S.}~\bibnamefont
  {Buchholz}},\ and\ \bibinfo {author} {\bibfnamefont {B.}~\bibnamefont
  {Sch{\"o}lkopf}},\ }\bibfield  {title} {\bibinfo {title} {The inductive bias
  of quantum kernels},\ }\href@noop {} {\bibfield  {journal} {\bibinfo
  {journal} {Advances in Neural Information Processing Systems}\ }\textbf
  {\bibinfo {volume} {34}} (\bibinfo {year} {2021})}\BibitemShut {NoStop}%
\bibitem [{\citenamefont {Wolpert}(1996)}]{wolpert1996lack}%
  \BibitemOpen
  \bibfield  {author} {\bibinfo {author} {\bibfnamefont {D.~H.}\ \bibnamefont
  {Wolpert}},\ }\bibfield  {title} {\bibinfo {title} {The lack of a priori
  distinctions between learning algorithms},\ }\href@noop {} {\bibfield
  {journal} {\bibinfo  {journal} {Neural computation}\ }\textbf {\bibinfo
  {volume} {8}},\ \bibinfo {pages} {1341} (\bibinfo {year} {1996})}\BibitemShut
  {NoStop}%
\bibitem [{\citenamefont {Paparo}\ \emph
  {et~al.}(2014{\natexlab{b}})\citenamefont {Paparo}, \citenamefont {Dunjko},
  \citenamefont {Makmal}, \citenamefont {Martin-Delgado},\ and\ \citenamefont
  {Briegel}}]{paparo2014quantum}%
  \BibitemOpen
  \bibfield  {author} {\bibinfo {author} {\bibfnamefont {G.~D.}\ \bibnamefont
  {Paparo}}, \bibinfo {author} {\bibfnamefont {V.}~\bibnamefont {Dunjko}},
  \bibinfo {author} {\bibfnamefont {A.}~\bibnamefont {Makmal}}, \bibinfo
  {author} {\bibfnamefont {M.~A.}\ \bibnamefont {Martin-Delgado}},\ and\
  \bibinfo {author} {\bibfnamefont {H.~J.}\ \bibnamefont {Briegel}},\
  }\bibfield  {title} {\bibinfo {title} {Quantum speedup for active learning
  agents},\ }\href@noop {} {\bibfield  {journal} {\bibinfo  {journal} {Physical
  Review X}\ }\textbf {\bibinfo {volume} {4}},\ \bibinfo {pages} {031002}
  (\bibinfo {year} {2014}{\natexlab{b}})}\BibitemShut {NoStop}%
\bibitem [{\citenamefont {Low}\ \emph {et~al.}(2014)\citenamefont {Low},
  \citenamefont {Yoder},\ and\ \citenamefont {Chuang}}]{low2014quantum}%
  \BibitemOpen
  \bibfield  {author} {\bibinfo {author} {\bibfnamefont {G.~H.}\ \bibnamefont
  {Low}}, \bibinfo {author} {\bibfnamefont {T.~J.}\ \bibnamefont {Yoder}},\
  and\ \bibinfo {author} {\bibfnamefont {I.~L.}\ \bibnamefont {Chuang}},\
  }\bibfield  {title} {\bibinfo {title} {Quantum inference on bayesian
  networks},\ }\href@noop {} {\bibfield  {journal} {\bibinfo  {journal}
  {Physical Review A}\ }\textbf {\bibinfo {volume} {89}},\ \bibinfo {pages}
  {062315} (\bibinfo {year} {2014})}\BibitemShut {NoStop}%
\bibitem [{\citenamefont {Babbush}\ \emph {et~al.}(2021)\citenamefont
  {Babbush}, \citenamefont {McClean}, \citenamefont {Newman}, \citenamefont
  {Gidney}, \citenamefont {Boixo},\ and\ \citenamefont {Neven}}]{babbush2021}%
  \BibitemOpen
  \bibfield  {author} {\bibinfo {author} {\bibfnamefont {R.}~\bibnamefont
  {Babbush}}, \bibinfo {author} {\bibfnamefont {J.~R.}\ \bibnamefont
  {McClean}}, \bibinfo {author} {\bibfnamefont {M.}~\bibnamefont {Newman}},
  \bibinfo {author} {\bibfnamefont {C.}~\bibnamefont {Gidney}}, \bibinfo
  {author} {\bibfnamefont {S.}~\bibnamefont {Boixo}},\ and\ \bibinfo {author}
  {\bibfnamefont {H.}~\bibnamefont {Neven}},\ }\bibfield  {title} {\bibinfo
  {title} {Focus beyond quadratic speedups for error-corrected quantum
  advantage},\ }\href {https://doi.org/10.1103/PRXQuantum.2.010103} {\bibfield
  {journal} {\bibinfo  {journal} {PRX Quantum}\ }\textbf {\bibinfo {volume}
  {2}},\ \bibinfo {pages} {010103} (\bibinfo {year} {2021})}\BibitemShut
  {NoStop}%
\bibitem [{\citenamefont {Aaronson}(2015)}]{aaronson15}%
  \BibitemOpen
  \bibfield  {author} {\bibinfo {author} {\bibfnamefont {S.}~\bibnamefont
  {Aaronson}},\ }\bibfield  {title} {\bibinfo {title} {Read the fine print},\
  }\href@noop {} {\bibfield  {journal} {\bibinfo  {journal} {Nature Physics}\
  }\textbf {\bibinfo {volume} {11}},\ \bibinfo {pages} {291} (\bibinfo {year}
  {2015})}\BibitemShut {NoStop}%
\bibitem [{\citenamefont {Tang}(2019)}]{tang2019quantum}%
  \BibitemOpen
  \bibfield  {author} {\bibinfo {author} {\bibfnamefont {E.}~\bibnamefont
  {Tang}},\ }\bibfield  {title} {\bibinfo {title} {A quantum-inspired classical
  algorithm for recommendation systems},\ }in\ \href@noop {} {\emph {\bibinfo
  {booktitle} {Proceedings of the 51st Annual ACM SIGACT Symposium on Theory of
  Computing}}}\ (\bibinfo {year} {2019})\ pp.\ \bibinfo {pages}
  {217--228}\BibitemShut {NoStop}%
\bibitem [{\citenamefont {Selig}\ \emph {et~al.}(2021)\citenamefont {Selig},
  \citenamefont {Murphy}, \citenamefont {Sundareswaran}, \citenamefont
  {Redmond},\ and\ \citenamefont {Caton}}]{selig2021case}%
  \BibitemOpen
  \bibfield  {author} {\bibinfo {author} {\bibfnamefont {P.}~\bibnamefont
  {Selig}}, \bibinfo {author} {\bibfnamefont {N.}~\bibnamefont {Murphy}},
  \bibinfo {author} {\bibfnamefont {A.}~\bibnamefont {Sundareswaran}}, \bibinfo
  {author} {\bibfnamefont {D.}~\bibnamefont {Redmond}},\ and\ \bibinfo {author}
  {\bibfnamefont {S.}~\bibnamefont {Caton}},\ }\bibfield  {title} {\bibinfo
  {title} {A case for noisy shallow gate-based circuits in quantum machine
  learning},\ }in\ \href@noop {} {\emph {\bibinfo {booktitle} {2021
  International Conference on Rebooting Computing (ICRC)}}}\ (\bibinfo
  {organization} {IEEE},\ \bibinfo {year} {2021})\ pp.\ \bibinfo {pages}
  {24--34}\BibitemShut {NoStop}%
\bibitem [{\citenamefont {Franz}\ \emph {et~al.}(2022)\citenamefont {Franz},
  \citenamefont {Wolf}, \citenamefont {Periyasamy}, \citenamefont {Ufrecht},
  \citenamefont {Scherer}, \citenamefont {Plinge}, \citenamefont {Mutschler},\
  and\ \citenamefont {Mauerer}}]{franz2022uncovering}%
  \BibitemOpen
  \bibfield  {author} {\bibinfo {author} {\bibfnamefont {M.}~\bibnamefont
  {Franz}}, \bibinfo {author} {\bibfnamefont {L.}~\bibnamefont {Wolf}},
  \bibinfo {author} {\bibfnamefont {M.}~\bibnamefont {Periyasamy}}, \bibinfo
  {author} {\bibfnamefont {C.}~\bibnamefont {Ufrecht}}, \bibinfo {author}
  {\bibfnamefont {D.~D.}\ \bibnamefont {Scherer}}, \bibinfo {author}
  {\bibfnamefont {A.}~\bibnamefont {Plinge}}, \bibinfo {author} {\bibfnamefont
  {C.}~\bibnamefont {Mutschler}},\ and\ \bibinfo {author} {\bibfnamefont
  {W.}~\bibnamefont {Mauerer}},\ }\bibfield  {title} {\bibinfo {title}
  {Uncovering instabilities in variational-quantum deep q-networks},\
  }\href@noop {} {\bibfield  {journal} {\bibinfo  {journal} {arXiv preprint
  arXiv:2202.05195}\ } (\bibinfo {year} {2022})}\BibitemShut {NoStop}%
\bibitem [{\citenamefont {Rosenblatt}(1958)}]{rosenblatt58}%
  \BibitemOpen
  \bibfield  {author} {\bibinfo {author} {\bibfnamefont {F.}~\bibnamefont
  {Rosenblatt}},\ }\bibfield  {title} {\bibinfo {title} {The perceptron: {A}
  probabilistic model for information storage and organization in the brain.},\
  }\href@noop {} {\bibfield  {journal} {\bibinfo  {journal} {Psychological
  Review}\ }\textbf {\bibinfo {volume} {65}},\ \bibinfo {pages} {386} (\bibinfo
  {year} {1958})}\BibitemShut {NoStop}%
\bibitem [{\citenamefont {Lewenstein}(1994)}]{lewenstein1994quantum}%
  \BibitemOpen
  \bibfield  {author} {\bibinfo {author} {\bibfnamefont {M.}~\bibnamefont
  {Lewenstein}},\ }\bibfield  {title} {\bibinfo {title} {Quantum perceptrons},\
  }\href@noop {} {\bibfield  {journal} {\bibinfo  {journal} {Journal of Modern
  Optics}\ }\textbf {\bibinfo {volume} {41}},\ \bibinfo {pages} {2491}
  (\bibinfo {year} {1994})}\BibitemShut {NoStop}%
\bibitem [{\citenamefont {Cao}\ \emph {et~al.}(2017)\citenamefont {Cao},
  \citenamefont {Guerreschi},\ and\ \citenamefont {Aspuru-Guzik}}]{cao17}%
  \BibitemOpen
  \bibfield  {author} {\bibinfo {author} {\bibfnamefont {Y.}~\bibnamefont
  {Cao}}, \bibinfo {author} {\bibfnamefont {G.~G.}\ \bibnamefont
  {Guerreschi}},\ and\ \bibinfo {author} {\bibfnamefont {A.}~\bibnamefont
  {Aspuru-Guzik}},\ }\bibfield  {title} {\bibinfo {title} {Quantum neuron: an
  elementary building block for machine learning on quantum computers},\
  }\href@noop {} {\bibfield  {journal} {\bibinfo  {journal} {arXiv preprint
  arXiv:1711.11240}\ } (\bibinfo {year} {2017})}\BibitemShut {NoStop}%
\bibitem [{\citenamefont {Beer}\ \emph {et~al.}(2020)\citenamefont {Beer},
  \citenamefont {Bondarenko}, \citenamefont {Farrelly}, \citenamefont
  {Osborne}, \citenamefont {Salzmann}, \citenamefont {Scheiermann},\ and\
  \citenamefont {Wolf}}]{beer2020training}%
  \BibitemOpen
  \bibfield  {author} {\bibinfo {author} {\bibfnamefont {K.}~\bibnamefont
  {Beer}}, \bibinfo {author} {\bibfnamefont {D.}~\bibnamefont {Bondarenko}},
  \bibinfo {author} {\bibfnamefont {T.}~\bibnamefont {Farrelly}}, \bibinfo
  {author} {\bibfnamefont {T.~J.}\ \bibnamefont {Osborne}}, \bibinfo {author}
  {\bibfnamefont {R.}~\bibnamefont {Salzmann}}, \bibinfo {author}
  {\bibfnamefont {D.}~\bibnamefont {Scheiermann}},\ and\ \bibinfo {author}
  {\bibfnamefont {R.}~\bibnamefont {Wolf}},\ }\bibfield  {title} {\bibinfo
  {title} {Training deep quantum neural networks},\ }\href@noop {} {\bibfield
  {journal} {\bibinfo  {journal} {Nature communications}\ }\textbf {\bibinfo
  {volume} {11}},\ \bibinfo {pages} {1} (\bibinfo {year} {2020})}\BibitemShut
  {NoStop}%
\bibitem [{\citenamefont {Tacchino}\ \emph {et~al.}(2019)\citenamefont
  {Tacchino}, \citenamefont {Macchiavello}, \citenamefont {Gerace},\ and\
  \citenamefont {Bajoni}}]{tacchino2019artificial}%
  \BibitemOpen
  \bibfield  {author} {\bibinfo {author} {\bibfnamefont {F.}~\bibnamefont
  {Tacchino}}, \bibinfo {author} {\bibfnamefont {C.}~\bibnamefont
  {Macchiavello}}, \bibinfo {author} {\bibfnamefont {D.}~\bibnamefont
  {Gerace}},\ and\ \bibinfo {author} {\bibfnamefont {D.}~\bibnamefont
  {Bajoni}},\ }\bibfield  {title} {\bibinfo {title} {An artificial neuron
  implemented on an actual quantum processor},\ }\href@noop {} {\bibfield
  {journal} {\bibinfo  {journal} {npj Quantum Information}\ }\textbf {\bibinfo
  {volume} {5}},\ \bibinfo {pages} {1} (\bibinfo {year} {2019})}\BibitemShut
  {NoStop}%
\bibitem [{\citenamefont {Taroni}(2015)}]{taroni201590}%
  \BibitemOpen
  \bibfield  {author} {\bibinfo {author} {\bibfnamefont {A.}~\bibnamefont
  {Taroni}},\ }\bibfield  {title} {\bibinfo {title} {90 years of the ising
  model},\ }\href@noop {} {\bibfield  {journal} {\bibinfo  {journal} {Nature
  Physics}\ }\textbf {\bibinfo {volume} {11}},\ \bibinfo {pages} {997}
  (\bibinfo {year} {2015})}\BibitemShut {NoStop}%
\bibitem [{\citenamefont {Bartlett}\ \emph {et~al.}(2020)\citenamefont
  {Bartlett}, \citenamefont {Long}, \citenamefont {Lugosi},\ and\ \citenamefont
  {Tsigler}}]{bartlett2020benign}%
  \BibitemOpen
  \bibfield  {author} {\bibinfo {author} {\bibfnamefont {P.~L.}\ \bibnamefont
  {Bartlett}}, \bibinfo {author} {\bibfnamefont {P.~M.}\ \bibnamefont {Long}},
  \bibinfo {author} {\bibfnamefont {G.}~\bibnamefont {Lugosi}},\ and\ \bibinfo
  {author} {\bibfnamefont {A.}~\bibnamefont {Tsigler}},\ }\bibfield  {title}
  {\bibinfo {title} {Benign overfitting in linear regression},\ }\href@noop {}
  {\bibfield  {journal} {\bibinfo  {journal} {Proceedings of the National
  Academy of Sciences}\ }\textbf {\bibinfo {volume} {117}},\ \bibinfo {pages}
  {30063} (\bibinfo {year} {2020})}\BibitemShut {NoStop}%
\bibitem [{\citenamefont {Saxe}\ \emph {et~al.}(2013)\citenamefont {Saxe},
  \citenamefont {McClelland},\ and\ \citenamefont {Ganguli}}]{saxe2013exact}%
  \BibitemOpen
  \bibfield  {author} {\bibinfo {author} {\bibfnamefont {A.~M.}\ \bibnamefont
  {Saxe}}, \bibinfo {author} {\bibfnamefont {J.~L.}\ \bibnamefont
  {McClelland}},\ and\ \bibinfo {author} {\bibfnamefont {S.}~\bibnamefont
  {Ganguli}},\ }\bibfield  {title} {\bibinfo {title} {Exact solutions to the
  nonlinear dynamics of learning in deep linear neural networks},\ }\href@noop
  {} {\bibfield  {journal} {\bibinfo  {journal} {arXiv preprint
  arXiv:1312.6120}\ } (\bibinfo {year} {2013})}\BibitemShut {NoStop}%
\bibitem [{\citenamefont {Chatterjee}\ and\ \citenamefont
  {Yu}(2016)}]{chatterjee2016generalized}%
  \BibitemOpen
  \bibfield  {author} {\bibinfo {author} {\bibfnamefont {R.}~\bibnamefont
  {Chatterjee}}\ and\ \bibinfo {author} {\bibfnamefont {T.}~\bibnamefont
  {Yu}},\ }\bibfield  {title} {\bibinfo {title} {Generalized coherent states,
  reproducing kernels, and quantum support vector machines},\ }\href@noop {}
  {\bibfield  {journal} {\bibinfo  {journal} {arXiv preprint arXiv:1612.03713}\
  } (\bibinfo {year} {2016})}\BibitemShut {NoStop}%
\bibitem [{\citenamefont {Schuld}\ and\ \citenamefont
  {Killoran}(2019)}]{schuld2019quantum}%
  \BibitemOpen
  \bibfield  {author} {\bibinfo {author} {\bibfnamefont {M.}~\bibnamefont
  {Schuld}}\ and\ \bibinfo {author} {\bibfnamefont {N.}~\bibnamefont
  {Killoran}},\ }\bibfield  {title} {\bibinfo {title} {Quantum machine learning
  in feature hilbert spaces},\ }\href@noop {} {\bibfield  {journal} {\bibinfo
  {journal} {Physical review letters}\ }\textbf {\bibinfo {volume} {122}},\
  \bibinfo {pages} {040504} (\bibinfo {year} {2019})}\BibitemShut {NoStop}%
\bibitem [{\citenamefont {Havl{\'\i}{\v{c}}ek}\ \emph
  {et~al.}(2019)\citenamefont {Havl{\'\i}{\v{c}}ek}, \citenamefont
  {C{\'o}rcoles}, \citenamefont {Temme}, \citenamefont {Harrow}, \citenamefont
  {Kandala}, \citenamefont {Chow},\ and\ \citenamefont
  {Gambetta}}]{havlivcek2019supervised}%
  \BibitemOpen
  \bibfield  {author} {\bibinfo {author} {\bibfnamefont {V.}~\bibnamefont
  {Havl{\'\i}{\v{c}}ek}}, \bibinfo {author} {\bibfnamefont {A.~D.}\
  \bibnamefont {C{\'o}rcoles}}, \bibinfo {author} {\bibfnamefont
  {K.}~\bibnamefont {Temme}}, \bibinfo {author} {\bibfnamefont {A.~W.}\
  \bibnamefont {Harrow}}, \bibinfo {author} {\bibfnamefont {A.}~\bibnamefont
  {Kandala}}, \bibinfo {author} {\bibfnamefont {J.~M.}\ \bibnamefont {Chow}},\
  and\ \bibinfo {author} {\bibfnamefont {J.~M.}\ \bibnamefont {Gambetta}},\
  }\bibfield  {title} {\bibinfo {title} {Supervised learning with
  quantum-enhanced feature spaces},\ }\href@noop {} {\bibfield  {journal}
  {\bibinfo  {journal} {Nature}\ }\textbf {\bibinfo {volume} {567}},\ \bibinfo
  {pages} {209} (\bibinfo {year} {2019})}\BibitemShut {NoStop}%
\bibitem [{\citenamefont {Schuld}(2021)}]{schuld2021quantum}%
  \BibitemOpen
  \bibfield  {author} {\bibinfo {author} {\bibfnamefont {M.}~\bibnamefont
  {Schuld}},\ }\bibfield  {title} {\bibinfo {title} {Supervised quantum machine
  learning models are kernel methods},\ }\href@noop {} {\bibfield  {journal}
  {\bibinfo  {journal} {arXiv preprint arXiv:2101.11020}\ } (\bibinfo {year}
  {2021})}\BibitemShut {NoStop}%
\bibitem [{\citenamefont {Gentinetta}\ \emph {et~al.}(2022)\citenamefont
  {Gentinetta}, \citenamefont {Thomsen}, \citenamefont {Sutter},\ and\
  \citenamefont {Woerner}}]{gentinetta2022complexity}%
  \BibitemOpen
  \bibfield  {author} {\bibinfo {author} {\bibfnamefont {G.}~\bibnamefont
  {Gentinetta}}, \bibinfo {author} {\bibfnamefont {A.}~\bibnamefont {Thomsen}},
  \bibinfo {author} {\bibfnamefont {D.}~\bibnamefont {Sutter}},\ and\ \bibinfo
  {author} {\bibfnamefont {S.}~\bibnamefont {Woerner}},\ }\bibfield  {title}
  {\bibinfo {title} {The complexity of quantum support vector machines},\
  }\href@noop {} {\bibfield  {journal} {\bibinfo  {journal} {arXiv preprint
  arXiv:2203.00031}\ } (\bibinfo {year} {2022})}\BibitemShut {NoStop}%
\bibitem [{\citenamefont {Peters}\ \emph {et~al.}(2021)\citenamefont {Peters},
  \citenamefont {Caldeira}, \citenamefont {Ho}, \citenamefont {Leichenauer},
  \citenamefont {Mohseni}, \citenamefont {Neven}, \citenamefont {Spentzouris},
  \citenamefont {Strain},\ and\ \citenamefont {Perdue}}]{peters2021machine}%
  \BibitemOpen
  \bibfield  {author} {\bibinfo {author} {\bibfnamefont {E.}~\bibnamefont
  {Peters}}, \bibinfo {author} {\bibfnamefont {J.}~\bibnamefont {Caldeira}},
  \bibinfo {author} {\bibfnamefont {A.}~\bibnamefont {Ho}}, \bibinfo {author}
  {\bibfnamefont {S.}~\bibnamefont {Leichenauer}}, \bibinfo {author}
  {\bibfnamefont {M.}~\bibnamefont {Mohseni}}, \bibinfo {author} {\bibfnamefont
  {H.}~\bibnamefont {Neven}}, \bibinfo {author} {\bibfnamefont
  {P.}~\bibnamefont {Spentzouris}}, \bibinfo {author} {\bibfnamefont
  {D.}~\bibnamefont {Strain}},\ and\ \bibinfo {author} {\bibfnamefont {G.~N.}\
  \bibnamefont {Perdue}},\ }\bibfield  {title} {\bibinfo {title} {Machine
  learning of high dimensional data on a noisy quantum processor},\ }\href@noop
  {} {\bibfield  {journal} {\bibinfo  {journal} {npj Quantum Information}\
  }\textbf {\bibinfo {volume} {7}},\ \bibinfo {pages} {1} (\bibinfo {year}
  {2021})}\BibitemShut {NoStop}%
\bibitem [{\citenamefont {Rahimi}\ \emph {et~al.}(2007)\citenamefont {Rahimi},
  \citenamefont {Recht} \emph {et~al.}}]{rahimi2007random}%
  \BibitemOpen
  \bibfield  {author} {\bibinfo {author} {\bibfnamefont {A.}~\bibnamefont
  {Rahimi}}, \bibinfo {author} {\bibfnamefont {B.}~\bibnamefont {Recht}}, \emph
  {et~al.},\ }\bibfield  {title} {\bibinfo {title} {Random features for
  large-scale kernel machines.},\ }in\ \href@noop {} {\emph {\bibinfo
  {booktitle} {NIPS}}},\ Vol.~\bibinfo {volume} {3}\ (\bibinfo {organization}
  {Citeseer},\ \bibinfo {year} {2007})\ p.~\bibinfo {pages} {5}\BibitemShut
  {NoStop}%
\bibitem [{\citenamefont {Jacot}\ \emph {et~al.}(2018)\citenamefont {Jacot},
  \citenamefont {Gabriel},\ and\ \citenamefont {Hongler}}]{jacot2018neural}%
  \BibitemOpen
  \bibfield  {author} {\bibinfo {author} {\bibfnamefont {A.}~\bibnamefont
  {Jacot}}, \bibinfo {author} {\bibfnamefont {F.}~\bibnamefont {Gabriel}},\
  and\ \bibinfo {author} {\bibfnamefont {C.}~\bibnamefont {Hongler}},\
  }\bibfield  {title} {\bibinfo {title} {Neural tangent kernel: Convergence and
  generalization in neural networks},\ }\href@noop {} {\bibfield  {journal}
  {\bibinfo  {journal} {arXiv preprint arXiv:1806.07572}\ } (\bibinfo {year}
  {2018})}\BibitemShut {NoStop}%
\bibitem [{\citenamefont {Shirai}\ \emph {et~al.}(2021)\citenamefont {Shirai},
  \citenamefont {Kubo}, \citenamefont {Mitarai},\ and\ \citenamefont
  {Fujii}}]{shirai2021quantum}%
  \BibitemOpen
  \bibfield  {author} {\bibinfo {author} {\bibfnamefont {N.}~\bibnamefont
  {Shirai}}, \bibinfo {author} {\bibfnamefont {K.}~\bibnamefont {Kubo}},
  \bibinfo {author} {\bibfnamefont {K.}~\bibnamefont {Mitarai}},\ and\ \bibinfo
  {author} {\bibfnamefont {K.}~\bibnamefont {Fujii}},\ }\bibfield  {title}
  {\bibinfo {title} {Quantum tangent kernel},\ }\href@noop {} {\bibfield
  {journal} {\bibinfo  {journal} {arXiv preprint arXiv:2111.02951}\ } (\bibinfo
  {year} {2021})}\BibitemShut {NoStop}%
\bibitem [{\citenamefont {Nakaji}\ \emph {et~al.}(2021)\citenamefont {Nakaji},
  \citenamefont {Tezuka},\ and\ \citenamefont {Yamamoto}}]{nakaji2021quantum}%
  \BibitemOpen
  \bibfield  {author} {\bibinfo {author} {\bibfnamefont {K.}~\bibnamefont
  {Nakaji}}, \bibinfo {author} {\bibfnamefont {H.}~\bibnamefont {Tezuka}},\
  and\ \bibinfo {author} {\bibfnamefont {N.}~\bibnamefont {Yamamoto}},\
  }\bibfield  {title} {\bibinfo {title} {Quantum-enhanced neural networks in
  the neural tangent kernel framework},\ }\href@noop {} {\bibfield  {journal}
  {\bibinfo  {journal} {arXiv preprint arXiv:2109.03786}\ } (\bibinfo {year}
  {2021})}\BibitemShut {NoStop}%
\bibitem [{\citenamefont {Liu}\ \emph {et~al.}(2021{\natexlab{b}})\citenamefont
  {Liu}, \citenamefont {Tacchino}, \citenamefont {Glick}, \citenamefont
  {Jiang},\ and\ \citenamefont {Mezzacapo}}]{liu2021representation}%
  \BibitemOpen
  \bibfield  {author} {\bibinfo {author} {\bibfnamefont {J.}~\bibnamefont
  {Liu}}, \bibinfo {author} {\bibfnamefont {F.}~\bibnamefont {Tacchino}},
  \bibinfo {author} {\bibfnamefont {J.~R.}\ \bibnamefont {Glick}}, \bibinfo
  {author} {\bibfnamefont {L.}~\bibnamefont {Jiang}},\ and\ \bibinfo {author}
  {\bibfnamefont {A.}~\bibnamefont {Mezzacapo}},\ }\bibfield  {title} {\bibinfo
  {title} {Representation learning via quantum neural tangent kernels},\
  }\href@noop {} {\bibfield  {journal} {\bibinfo  {journal} {arXiv preprint
  arXiv:2111.04225}\ } (\bibinfo {year} {2021}{\natexlab{b}})}\BibitemShut
  {NoStop}%
\bibitem [{\citenamefont {Steinwart}\ and\ \citenamefont
  {Christmann}(2008)}]{steinwart2008}%
  \BibitemOpen
  \bibfield  {author} {\bibinfo {author} {\bibfnamefont {I.}~\bibnamefont
  {Steinwart}}\ and\ \bibinfo {author} {\bibfnamefont {A.}~\bibnamefont
  {Christmann}},\ }\href@noop {} {\emph {\bibinfo {title} {Support Vector
  Machines}}},\ \bibinfo {edition} {1st}\ ed.\ (\bibinfo  {publisher} {Springer
  Publishing Company, Incorporated},\ \bibinfo {year} {2008})\BibitemShut
  {NoStop}%
\bibitem [{\citenamefont {Lloyd}\ \emph {et~al.}(2020)\citenamefont {Lloyd},
  \citenamefont {Schuld}, \citenamefont {Ijaz}, \citenamefont {Izaac},\ and\
  \citenamefont {Killoran}}]{lloyd2020quantum}%
  \BibitemOpen
  \bibfield  {author} {\bibinfo {author} {\bibfnamefont {S.}~\bibnamefont
  {Lloyd}}, \bibinfo {author} {\bibfnamefont {M.}~\bibnamefont {Schuld}},
  \bibinfo {author} {\bibfnamefont {A.}~\bibnamefont {Ijaz}}, \bibinfo {author}
  {\bibfnamefont {J.}~\bibnamefont {Izaac}},\ and\ \bibinfo {author}
  {\bibfnamefont {N.}~\bibnamefont {Killoran}},\ }\bibfield  {title} {\bibinfo
  {title} {Quantum embeddings for machine learning},\ }\href@noop {} {\bibfield
   {journal} {\bibinfo  {journal} {arXiv preprint arXiv:2001.03622}\ }
  (\bibinfo {year} {2020})}\BibitemShut {NoStop}%
\bibitem [{\citenamefont {Cheng}\ \emph {et~al.}(2018)\citenamefont {Cheng},
  \citenamefont {Chen},\ and\ \citenamefont {Wang}}]{cheng2018information}%
  \BibitemOpen
  \bibfield  {author} {\bibinfo {author} {\bibfnamefont {S.}~\bibnamefont
  {Cheng}}, \bibinfo {author} {\bibfnamefont {J.}~\bibnamefont {Chen}},\ and\
  \bibinfo {author} {\bibfnamefont {L.}~\bibnamefont {Wang}},\ }\bibfield
  {title} {\bibinfo {title} {Information perspective to probabilistic modeling:
  Boltzmann machines versus born machines},\ }\href@noop {} {\bibfield
  {journal} {\bibinfo  {journal} {Entropy}\ }\textbf {\bibinfo {volume} {20}},\
  \bibinfo {pages} {583} (\bibinfo {year} {2018})}\BibitemShut {NoStop}%
\bibitem [{\citenamefont {Carleo}\ and\ \citenamefont
  {Troyer}(2017)}]{carleo2017solving}%
  \BibitemOpen
  \bibfield  {author} {\bibinfo {author} {\bibfnamefont {G.}~\bibnamefont
  {Carleo}}\ and\ \bibinfo {author} {\bibfnamefont {M.}~\bibnamefont
  {Troyer}},\ }\bibfield  {title} {\bibinfo {title} {Solving the quantum
  many-body problem with artificial neural networks},\ }\href@noop {}
  {\bibfield  {journal} {\bibinfo  {journal} {Science}\ }\textbf {\bibinfo
  {volume} {355}},\ \bibinfo {pages} {602} (\bibinfo {year}
  {2017})}\BibitemShut {NoStop}%
\bibitem [{\citenamefont {Cerezo}\ \emph {et~al.}(2021)\citenamefont {Cerezo},
  \citenamefont {Arrasmith}, \citenamefont {Babbush}, \citenamefont {Benjamin},
  \citenamefont {Endo}, \citenamefont {Fujii}, \citenamefont {McClean},
  \citenamefont {Mitarai}, \citenamefont {Yuan}, \citenamefont {Cincio} \emph
  {et~al.}}]{cerezo2021variational}%
  \BibitemOpen
  \bibfield  {author} {\bibinfo {author} {\bibfnamefont {M.}~\bibnamefont
  {Cerezo}}, \bibinfo {author} {\bibfnamefont {A.}~\bibnamefont {Arrasmith}},
  \bibinfo {author} {\bibfnamefont {R.}~\bibnamefont {Babbush}}, \bibinfo
  {author} {\bibfnamefont {S.~C.}\ \bibnamefont {Benjamin}}, \bibinfo {author}
  {\bibfnamefont {S.}~\bibnamefont {Endo}}, \bibinfo {author} {\bibfnamefont
  {K.}~\bibnamefont {Fujii}}, \bibinfo {author} {\bibfnamefont {J.~R.}\
  \bibnamefont {McClean}}, \bibinfo {author} {\bibfnamefont {K.}~\bibnamefont
  {Mitarai}}, \bibinfo {author} {\bibfnamefont {X.}~\bibnamefont {Yuan}},
  \bibinfo {author} {\bibfnamefont {L.}~\bibnamefont {Cincio}}, \emph
  {et~al.},\ }\bibfield  {title} {\bibinfo {title} {Variational quantum
  algorithms},\ }\href@noop {} {\bibfield  {journal} {\bibinfo  {journal}
  {Nature Reviews Physics}\ }\textbf {\bibinfo {volume} {3}},\ \bibinfo {pages}
  {625} (\bibinfo {year} {2021})}\BibitemShut {NoStop}%
\bibitem [{\citenamefont {Abadi}\ \emph {et~al.}(2016)\citenamefont {Abadi},
  \citenamefont {Barham}, \citenamefont {Chen}, \citenamefont {Chen},
  \citenamefont {Davis}, \citenamefont {Dean}, \citenamefont {Devin},
  \citenamefont {Ghemawat}, \citenamefont {Irving}, \citenamefont {Isard} \emph
  {et~al.}}]{abadi2016tensorflow}%
  \BibitemOpen
  \bibfield  {author} {\bibinfo {author} {\bibfnamefont {M.}~\bibnamefont
  {Abadi}}, \bibinfo {author} {\bibfnamefont {P.}~\bibnamefont {Barham}},
  \bibinfo {author} {\bibfnamefont {J.}~\bibnamefont {Chen}}, \bibinfo {author}
  {\bibfnamefont {Z.}~\bibnamefont {Chen}}, \bibinfo {author} {\bibfnamefont
  {A.}~\bibnamefont {Davis}}, \bibinfo {author} {\bibfnamefont
  {J.}~\bibnamefont {Dean}}, \bibinfo {author} {\bibfnamefont {M.}~\bibnamefont
  {Devin}}, \bibinfo {author} {\bibfnamefont {S.}~\bibnamefont {Ghemawat}},
  \bibinfo {author} {\bibfnamefont {G.}~\bibnamefont {Irving}}, \bibinfo
  {author} {\bibfnamefont {M.}~\bibnamefont {Isard}}, \emph {et~al.},\
  }\bibfield  {title} {\bibinfo {title} {Tensorflow: A system for large-scale
  machine learning},\ }in\ \href@noop {} {\emph {\bibinfo {booktitle} {12th
  $\{$USENIX$\}$ symposium on operating systems design and implementation
  ($\{$OSDI$\}$ 16)}}}\ (\bibinfo {year} {2016})\ pp.\ \bibinfo {pages}
  {265--283}\BibitemShut {NoStop}%
\bibitem [{\citenamefont {Paszke}\ \emph {et~al.}(2019)\citenamefont {Paszke},
  \citenamefont {Gross}, \citenamefont {Massa}, \citenamefont {Lerer},
  \citenamefont {Bradbury}, \citenamefont {Chanan}, \citenamefont {Killeen},
  \citenamefont {Lin}, \citenamefont {Gimelshein}, \citenamefont {Antiga} \emph
  {et~al.}}]{paszke2019pytorch}%
  \BibitemOpen
  \bibfield  {author} {\bibinfo {author} {\bibfnamefont {A.}~\bibnamefont
  {Paszke}}, \bibinfo {author} {\bibfnamefont {S.}~\bibnamefont {Gross}},
  \bibinfo {author} {\bibfnamefont {F.}~\bibnamefont {Massa}}, \bibinfo
  {author} {\bibfnamefont {A.}~\bibnamefont {Lerer}}, \bibinfo {author}
  {\bibfnamefont {J.}~\bibnamefont {Bradbury}}, \bibinfo {author}
  {\bibfnamefont {G.}~\bibnamefont {Chanan}}, \bibinfo {author} {\bibfnamefont
  {T.}~\bibnamefont {Killeen}}, \bibinfo {author} {\bibfnamefont
  {Z.}~\bibnamefont {Lin}}, \bibinfo {author} {\bibfnamefont {N.}~\bibnamefont
  {Gimelshein}}, \bibinfo {author} {\bibfnamefont {L.}~\bibnamefont {Antiga}},
  \emph {et~al.},\ }\bibfield  {title} {\bibinfo {title} {Pytorch: An
  imperative style, high-performance deep learning library},\ }\href@noop {}
  {\bibfield  {journal} {\bibinfo  {journal} {Advances in neural information
  processing systems}\ }\textbf {\bibinfo {volume} {32}},\ \bibinfo {pages}
  {8026} (\bibinfo {year} {2019})}\BibitemShut {NoStop}%
\bibitem [{\citenamefont {Rumelhart}\ \emph {et~al.}(1986)\citenamefont
  {Rumelhart}, \citenamefont {Hinton},\ and\ \citenamefont
  {Williams}}]{rumelhart1986learning}%
  \BibitemOpen
  \bibfield  {author} {\bibinfo {author} {\bibfnamefont {D.~E.}\ \bibnamefont
  {Rumelhart}}, \bibinfo {author} {\bibfnamefont {G.~E.}\ \bibnamefont
  {Hinton}},\ and\ \bibinfo {author} {\bibfnamefont {R.~J.}\ \bibnamefont
  {Williams}},\ }\bibfield  {title} {\bibinfo {title} {Learning representations
  by back-propagating errors},\ }\href@noop {} {\bibfield  {journal} {\bibinfo
  {journal} {nature}\ }\textbf {\bibinfo {volume} {323}},\ \bibinfo {pages}
  {533} (\bibinfo {year} {1986})}\BibitemShut {NoStop}%
\bibitem [{\citenamefont {Li}\ \emph {et~al.}(2017)\citenamefont {Li},
  \citenamefont {Yang}, \citenamefont {Peng},\ and\ \citenamefont
  {Sun}}]{li2017hybrid}%
  \BibitemOpen
  \bibfield  {author} {\bibinfo {author} {\bibfnamefont {J.}~\bibnamefont
  {Li}}, \bibinfo {author} {\bibfnamefont {X.}~\bibnamefont {Yang}}, \bibinfo
  {author} {\bibfnamefont {X.}~\bibnamefont {Peng}},\ and\ \bibinfo {author}
  {\bibfnamefont {C.-P.}\ \bibnamefont {Sun}},\ }\bibfield  {title} {\bibinfo
  {title} {Hybrid quantum-classical approach to quantum optimal control},\
  }\href@noop {} {\bibfield  {journal} {\bibinfo  {journal} {Physical review
  letters}\ }\textbf {\bibinfo {volume} {118}},\ \bibinfo {pages} {150503}
  (\bibinfo {year} {2017})}\BibitemShut {NoStop}%
\bibitem [{\citenamefont {Crooks}(2019)}]{crooks2019gradients}%
  \BibitemOpen
  \bibfield  {author} {\bibinfo {author} {\bibfnamefont {G.~E.}\ \bibnamefont
  {Crooks}},\ }\bibfield  {title} {\bibinfo {title} {Gradients of parameterized
  quantum gates using the parameter-shift rule and gate decomposition},\
  }\href@noop {} {\bibfield  {journal} {\bibinfo  {journal} {arXiv preprint
  arXiv:1905.13311}\ } (\bibinfo {year} {2019})}\BibitemShut {NoStop}%
\bibitem [{\citenamefont {Banchi}\ and\ \citenamefont
  {Crooks}(2021)}]{banchi2021measuring}%
  \BibitemOpen
  \bibfield  {author} {\bibinfo {author} {\bibfnamefont {L.}~\bibnamefont
  {Banchi}}\ and\ \bibinfo {author} {\bibfnamefont {G.~E.}\ \bibnamefont
  {Crooks}},\ }\bibfield  {title} {\bibinfo {title} {Measuring analytic
  gradients of general quantum evolution with the stochastic parameter shift
  rule},\ }\href@noop {} {\bibfield  {journal} {\bibinfo  {journal} {Quantum}\
  }\textbf {\bibinfo {volume} {5}},\ \bibinfo {pages} {386} (\bibinfo {year}
  {2021})}\BibitemShut {NoStop}%
\bibitem [{\citenamefont {Kottmann}\ \emph {et~al.}(2021)\citenamefont
  {Kottmann}, \citenamefont {Anand},\ and\ \citenamefont
  {Aspuru-Guzik}}]{kottmann2021feasible}%
  \BibitemOpen
  \bibfield  {author} {\bibinfo {author} {\bibfnamefont {J.~S.}\ \bibnamefont
  {Kottmann}}, \bibinfo {author} {\bibfnamefont {A.}~\bibnamefont {Anand}},\
  and\ \bibinfo {author} {\bibfnamefont {A.}~\bibnamefont {Aspuru-Guzik}},\
  }\bibfield  {title} {\bibinfo {title} {A feasible approach for automatically
  differentiable unitary coupled-cluster on quantum computers},\ }\href@noop {}
  {\bibfield  {journal} {\bibinfo  {journal} {Chemical Science}\ }\textbf
  {\bibinfo {volume} {12}},\ \bibinfo {pages} {3497} (\bibinfo {year}
  {2021})}\BibitemShut {NoStop}%
\bibitem [{\citenamefont {Vidal}\ and\ \citenamefont
  {Theis}(2018)}]{vidal2018calculus}%
  \BibitemOpen
  \bibfield  {author} {\bibinfo {author} {\bibfnamefont {J.~G.}\ \bibnamefont
  {Vidal}}\ and\ \bibinfo {author} {\bibfnamefont {D.~O.}\ \bibnamefont
  {Theis}},\ }\bibfield  {title} {\bibinfo {title} {Calculus on parameterized
  quantum circuits},\ }\href@noop {} {\bibfield  {journal} {\bibinfo  {journal}
  {arXiv preprint arXiv:1812.06323}\ } (\bibinfo {year} {2018})}\BibitemShut
  {NoStop}%
\bibitem [{\citenamefont {Mari}\ \emph {et~al.}(2021)\citenamefont {Mari},
  \citenamefont {Bromley},\ and\ \citenamefont
  {Killoran}}]{mari2021estimating}%
  \BibitemOpen
  \bibfield  {author} {\bibinfo {author} {\bibfnamefont {A.}~\bibnamefont
  {Mari}}, \bibinfo {author} {\bibfnamefont {T.~R.}\ \bibnamefont {Bromley}},\
  and\ \bibinfo {author} {\bibfnamefont {N.}~\bibnamefont {Killoran}},\
  }\bibfield  {title} {\bibinfo {title} {Estimating the gradient and
  higher-order derivatives on quantum hardware},\ }\href@noop {} {\bibfield
  {journal} {\bibinfo  {journal} {Physical Review A}\ }\textbf {\bibinfo
  {volume} {103}},\ \bibinfo {pages} {012405} (\bibinfo {year}
  {2021})}\BibitemShut {NoStop}%
\bibitem [{\citenamefont {Izmaylov}\ \emph {et~al.}(2021)\citenamefont
  {Izmaylov}, \citenamefont {Lang},\ and\ \citenamefont
  {Yen}}]{izmaylov2021analytic}%
  \BibitemOpen
  \bibfield  {author} {\bibinfo {author} {\bibfnamefont {A.~F.}\ \bibnamefont
  {Izmaylov}}, \bibinfo {author} {\bibfnamefont {R.~A.}\ \bibnamefont {Lang}},\
  and\ \bibinfo {author} {\bibfnamefont {T.-C.}\ \bibnamefont {Yen}},\
  }\bibfield  {title} {\bibinfo {title} {Analytic gradients in variational
  quantum algorithms: Algebraic extensions of the parameter-shift rule to
  general unitary transformations},\ }\href@noop {} {\bibfield  {journal}
  {\bibinfo  {journal} {arXiv preprint arXiv:2107.08131}\ } (\bibinfo {year}
  {2021})}\BibitemShut {NoStop}%
\bibitem [{\citenamefont {Kyriienko}\ and\ \citenamefont
  {Elfving}(2021)}]{kyriienko2021generalized}%
  \BibitemOpen
  \bibfield  {author} {\bibinfo {author} {\bibfnamefont {O.}~\bibnamefont
  {Kyriienko}}\ and\ \bibinfo {author} {\bibfnamefont {V.~E.}\ \bibnamefont
  {Elfving}},\ }\bibfield  {title} {\bibinfo {title} {Generalized quantum
  circuit differentiation rules},\ }\href@noop {} {\bibfield  {journal}
  {\bibinfo  {journal} {arXiv preprint arXiv:2108.01218}\ } (\bibinfo {year}
  {2021})}\BibitemShut {NoStop}%
\bibitem [{\citenamefont {Wierichs}\ \emph {et~al.}(2021)\citenamefont
  {Wierichs}, \citenamefont {Izaac}, \citenamefont {Wang},\ and\ \citenamefont
  {Lin}}]{wierichs2021general}%
  \BibitemOpen
  \bibfield  {author} {\bibinfo {author} {\bibfnamefont {D.}~\bibnamefont
  {Wierichs}}, \bibinfo {author} {\bibfnamefont {J.}~\bibnamefont {Izaac}},
  \bibinfo {author} {\bibfnamefont {C.}~\bibnamefont {Wang}},\ and\ \bibinfo
  {author} {\bibfnamefont {C.~Y.-Y.}\ \bibnamefont {Lin}},\ }\bibfield  {title}
  {\bibinfo {title} {General parameter-shift rules for quantum gradients},\
  }\href@noop {} {\bibfield  {journal} {\bibinfo  {journal} {arXiv preprint
  arXiv:2107.12390}\ } (\bibinfo {year} {2021})}\BibitemShut {NoStop}%
\bibitem [{\citenamefont {Kingma}\ and\ \citenamefont
  {Ba}(2014)}]{kingma2014adam}%
  \BibitemOpen
  \bibfield  {author} {\bibinfo {author} {\bibfnamefont {D.~P.}\ \bibnamefont
  {Kingma}}\ and\ \bibinfo {author} {\bibfnamefont {J.}~\bibnamefont {Ba}},\
  }\bibfield  {title} {\bibinfo {title} {Adam: A method for stochastic
  optimization},\ }\href@noop {} {\bibfield  {journal} {\bibinfo  {journal}
  {arXiv preprint arXiv:1412.6980}\ } (\bibinfo {year} {2014})}\BibitemShut
  {NoStop}%
\bibitem [{\citenamefont {Stokes}\ \emph {et~al.}(2020)\citenamefont {Stokes},
  \citenamefont {Izaac}, \citenamefont {Killoran},\ and\ \citenamefont
  {Carleo}}]{stokes2020quantum}%
  \BibitemOpen
  \bibfield  {author} {\bibinfo {author} {\bibfnamefont {J.}~\bibnamefont
  {Stokes}}, \bibinfo {author} {\bibfnamefont {J.}~\bibnamefont {Izaac}},
  \bibinfo {author} {\bibfnamefont {N.}~\bibnamefont {Killoran}},\ and\
  \bibinfo {author} {\bibfnamefont {G.}~\bibnamefont {Carleo}},\ }\bibfield
  {title} {\bibinfo {title} {Quantum natural gradient},\ }\href@noop {}
  {\bibfield  {journal} {\bibinfo  {journal} {Quantum}\ }\textbf {\bibinfo
  {volume} {4}},\ \bibinfo {pages} {269} (\bibinfo {year} {2020})}\BibitemShut
  {NoStop}%
\bibitem [{\citenamefont {K{\"u}bler}\ \emph {et~al.}(2020)\citenamefont
  {K{\"u}bler}, \citenamefont {Arrasmith}, \citenamefont {Cincio},\ and\
  \citenamefont {Coles}}]{kubler2020adaptive}%
  \BibitemOpen
  \bibfield  {author} {\bibinfo {author} {\bibfnamefont {J.~M.}\ \bibnamefont
  {K{\"u}bler}}, \bibinfo {author} {\bibfnamefont {A.}~\bibnamefont
  {Arrasmith}}, \bibinfo {author} {\bibfnamefont {L.}~\bibnamefont {Cincio}},\
  and\ \bibinfo {author} {\bibfnamefont {P.~J.}\ \bibnamefont {Coles}},\
  }\bibfield  {title} {\bibinfo {title} {An adaptive optimizer for
  measurement-frugal variational algorithms},\ }\href@noop {} {\bibfield
  {journal} {\bibinfo  {journal} {Quantum}\ }\textbf {\bibinfo {volume} {4}},\
  \bibinfo {pages} {263} (\bibinfo {year} {2020})}\BibitemShut {NoStop}%
\bibitem [{\citenamefont {Arrasmith}\ \emph {et~al.}(2020)\citenamefont
  {Arrasmith}, \citenamefont {Cincio}, \citenamefont {Somma},\ and\
  \citenamefont {Coles}}]{arrasmith2020operator}%
  \BibitemOpen
  \bibfield  {author} {\bibinfo {author} {\bibfnamefont {A.}~\bibnamefont
  {Arrasmith}}, \bibinfo {author} {\bibfnamefont {L.}~\bibnamefont {Cincio}},
  \bibinfo {author} {\bibfnamefont {R.~D.}\ \bibnamefont {Somma}},\ and\
  \bibinfo {author} {\bibfnamefont {P.~J.}\ \bibnamefont {Coles}},\ }\bibfield
  {title} {\bibinfo {title} {Operator sampling for shot-frugal optimization in
  variational algorithms},\ }\href@noop {} {\bibfield  {journal} {\bibinfo
  {journal} {arXiv preprint arXiv:2004.06252}\ } (\bibinfo {year}
  {2020})}\BibitemShut {NoStop}%
\bibitem [{\citenamefont {Ostaszewski}\ \emph {et~al.}(2021)\citenamefont
  {Ostaszewski}, \citenamefont {Grant},\ and\ \citenamefont
  {Benedetti}}]{ostaszewski2021structure}%
  \BibitemOpen
  \bibfield  {author} {\bibinfo {author} {\bibfnamefont {M.}~\bibnamefont
  {Ostaszewski}}, \bibinfo {author} {\bibfnamefont {E.}~\bibnamefont {Grant}},\
  and\ \bibinfo {author} {\bibfnamefont {M.}~\bibnamefont {Benedetti}},\
  }\bibfield  {title} {\bibinfo {title} {Structure optimization for
  parameterized quantum circuits},\ }\href@noop {} {\bibfield  {journal}
  {\bibinfo  {journal} {Quantum}\ }\textbf {\bibinfo {volume} {5}},\ \bibinfo
  {pages} {391} (\bibinfo {year} {2021})}\BibitemShut {NoStop}%
\bibitem [{\citenamefont {Stoudenmire}(2018)}]{stoudenmire2018learning}%
  \BibitemOpen
  \bibfield  {author} {\bibinfo {author} {\bibfnamefont {E.~M.}\ \bibnamefont
  {Stoudenmire}},\ }\bibfield  {title} {\bibinfo {title} {Learning relevant
  features of data with multi-scale tensor networks},\ }\href@noop {}
  {\bibfield  {journal} {\bibinfo  {journal} {Quantum Science and Technology}\
  }\textbf {\bibinfo {volume} {3}},\ \bibinfo {pages} {034003} (\bibinfo {year}
  {2018})}\BibitemShut {NoStop}%
\end{thebibliography}%

\end{document}